\documentclass{elsart}

\usepackage{natbib}
\usepackage{hyperref}

\usepackage{graphicx}
\usepackage{amssymb}
\usepackage{amsmath}
\usepackage{marvosym}

\begin{document}

\begin{frontmatter}



  \title{All you ever wanted to know about optical long baseline
    stellar interferometry, but were too shy to ask your adviser}


  \author[MPIFR]{Florentin Millour}

  \address[MPIFR]{Max-Planck Institute F\"ur Radio-astronomie, Auf dem
    H\"ugel 69, 53121 Bonn, Germany}

  \begin{abstract}
    % Text of abstract
    I try to present a small view of the properties and issues
    related to astronomical interferometry observations. I
    recall a bit of history of the technique, give some basic assessments
    to the principle of interferometry, and finally, describe physical 
    processes and limitations that affect optical long baseline
    interferometry and which are, in general, very useful for everyday
    work. Therefore, this text is not intended to perform strong
    demonstrations and show accurate results, but rather to
    transmit the general ``feeling'' one needs to have to not be destabilised
    by the first contact to real world interferometry.
  \end{abstract}


\end{frontmatter}

% main text
\section{What is optical / IR long baseline interferometry ?}
\label{section1}

\subsection{In the beginning ... }

The discovery of the wave-property of light was probably made by
\citet{1800ooea.book.....Y}, who managed to produce light
interferences by letting it go through 2 close holes. Soon after this
discovery, 
\citet{Fizeau1851} proposed and \citet{Stephan1874} tried
unsuccessfully to use this wave-property of light to measure the
apparent diameter of stars at the \textit{Observatoire de Marseille}
with a 80 cm telescope.

\begin{figure}[htbp]
  \centering
  \begin{tabular}{cc}
    \includegraphics[width=0.4\textwidth]{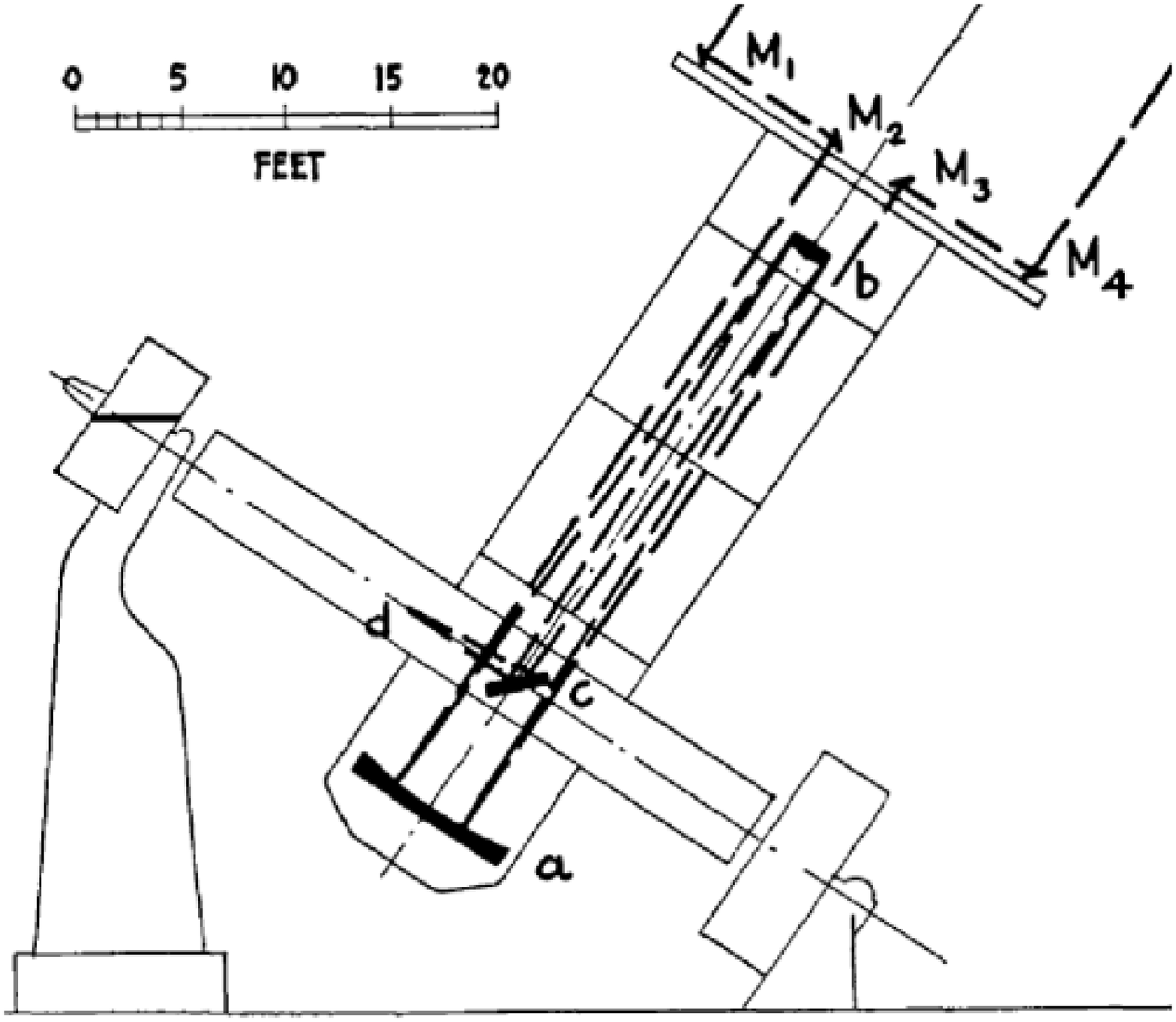}&
    \includegraphics[width=0.55\textwidth]{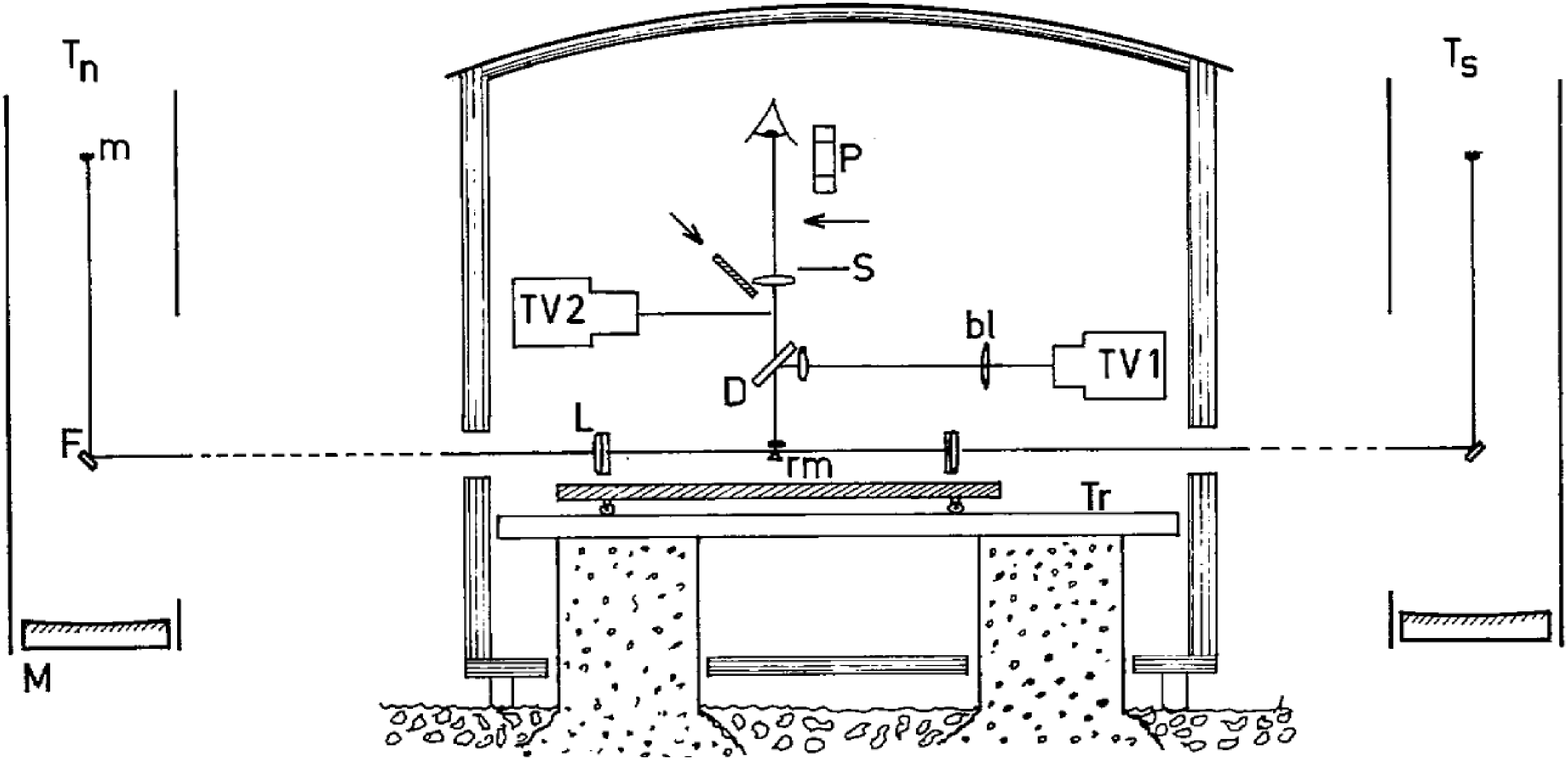}
  \end{tabular}
  \caption[The Michelson and the Labeyrie interferometers.]{
    \footnotesize{
      \emph{\bf Left:} The \emph{interferometer}
      of Michelson on the 100-inch telescope on the Mount Wilson
      \citep[taken from reference][]{1921ApJ....53..249M}.
      \emph{\bf Right:} The first \citet{1975ApJ...196L..71L}
      interferometer (diagram is from the article).
    }
  }
  \label{fig:Interferometre_Michelson}
\end{figure}

% \citet{1891PASP....3..274M} used this method to measure for the first
% time and with a good accuracy the diameters of the Jupiter satellites
% with the 100 inches ($\approx$ 2.50 m) telescope of the Mount
% Wilson. 
Then, \citet{1921ApJ....53..249M} managed to measure the
diameter of a star, Betelgeuse, equal to 0,047'' with a relative
accuracy of 10\% using a larger interferometer (see
Fig.~\ref{fig:Interferometre_Michelson}, left)
This experiment was very hard to carry out, the first results
\citep[][]{1921PASP...33..171P, 1922CMWCI.240....1M} were
so encouraging that a larger and more sensitive stellar
interferometer \citep{Pease1931} was built. However, technical
drawbacks and mechanical instabilities were the major problems of
these interferometers. Then, the project was abandonned during the
second world war.

When the very first radio-telescopes put into operation,
\citep{1947Obs....67...15R}, the idea to coherently combine together
several antennas was used to build radio-interferometers
\citep{1952RSPSA.211..351R}. This enabled an increase in the
spatial resolution of radio-collectors
\citep{1952PPSB...65..971S}. An easy access to the measures %  than in
% optical (one has access directly to the amplitude and phase of the
% electromagnetic field)
allowed a swift progress of the
instruments. This led to the VLA \citep[Very Large Array,
][]{2006IAUSS...1E..18B} and to the VLBA
\citep[Very Long Baselines Array, ][]{1975ApJ...201..249C}.

\vspace{1cm}
\vspace{1cm}

\subsection{Given up and reborn ...}

In the optical domain, the techniques developed by Michelson
were forgotten for more than 30 years. %because of higher difficulty
% in measuring the interference pattern than in radio-interferometry.
During this period, only intensity interferometry
\citep{1956Natur.178.1046H}, which indirectly recombined the light,
was developped and used for astronomy.

The true reborn of optical interferometry occurred in 1975 when
A. \citet{1975ApJ...196L..71L} managed to produce interference
fringes on a star using two separate telescopes (see
Fig.~\ref{fig:Interferometre_Michelson}, right).

This achievement was repeated in numerous institutes
where many prototypes were built and then used for science  (see
Table~\ref{tab:interferometres1}). The initial concept was improved
with the use of delay lines and stationary telescopes
\citep{1986MitAG..67..202M} rather than moving optical tables and
telescopes. The advent of \emph{spatial filtering}
\citep{1997ioai.book..115C} and \emph{simultaneous photometric
  calibration} allowed one to get rid of systematic effects
\citep{2003A&A...400.1173P, 2003A&A...398..385P}. This enabled an
advancement in accurate measurements.

\vspace{1cm}
\begin{table}[htbp]
  \caption[Interferometers in the world]{
    \footnotesize{
      Working interferometers in the world in alphabetical order. NIR
      means \emph{Near Infra-Red} (1-2.5$\mu$m) and MIR means
      \emph{Mid Infra-Red} (8-13$\mu$m).
    }
  }
  \begin{center}
    \begin{tabular}{|c|c|c|c|c|c|}
      \hline
      &\multicolumn{3}{c|}{Telescopes}&&\\
      Name & Diameter (m)& \# Combined & Total & Max. baseline & $\lambda$\\
      \hline
      \hline
      CHARA&1&2&6&330&Visible, NIR\\
      \hline
      COAST&0.4&3&6&47&Visible, NIR\\
      \hline
      % GI2T&1.52&2&2&65&Visible, NIR\\
      % \hline
      % IOTA&0.45&3&3&38&Visible, NIR\\
      % \hline
      ISI&1.65&3&3&85&MIR\\
      \hline
      Keck-I&10&2&2&85&NIR, MIR\\
      \hline
      MIRA-I&0.25&2&2&30&Visible\\
      \hline
      NPOI&0.12&6&6&64&Visible\\
      \hline
      PTI&0.4&3&3&110&NIR\\
      \hline
      SUSI&0.14&2&2&640&Visible\\
      \hline
      VLTI&8 / 1.8&3 / 3&4 / 4&130 / 200&NIR, MIR\\
      \hline
    \end{tabular}
  \end{center}
  \label{tab:interferometres1}
\end{table}
\vspace{1cm}

\subsection{Interferometry today}

Today's interferometry follows two different tracks: larger baselines
to get higher angular resolutions and larger telescope diameters to
reach higher magnitudes. The Keck-I and the VLTI, which combines both
large baselines and large apertures, are leading the way.

\paragraph*{Keck-I:}

The Keck-I is a US project consisting of two very large
10\,m segmented telescopes separated by 85 m.
The Keck telescopes work in two main modes: independently or combined
in the interferometric
mode. This last mode uses two instruments: a coaxial
re-combiner and a 2-telescope nuller. Today's K-band limiting
magnitude is 10, and it can work in low (R $\approx$ 20) and medium (R
$\approx$ 200) spectral resolutions. A differential mode and a phase
referencing mode are foreseen for the future.
However, the initially planned additional small telescopes for the
interferometers (the ``outriggers'') have been abandoned, limiting the
Keck-I project to a 2 telescope only interferometer.

\paragraph*{VLTI:}

\begin{figure}[htbp]
  \centering
  \includegraphics[width=0.95\textwidth]{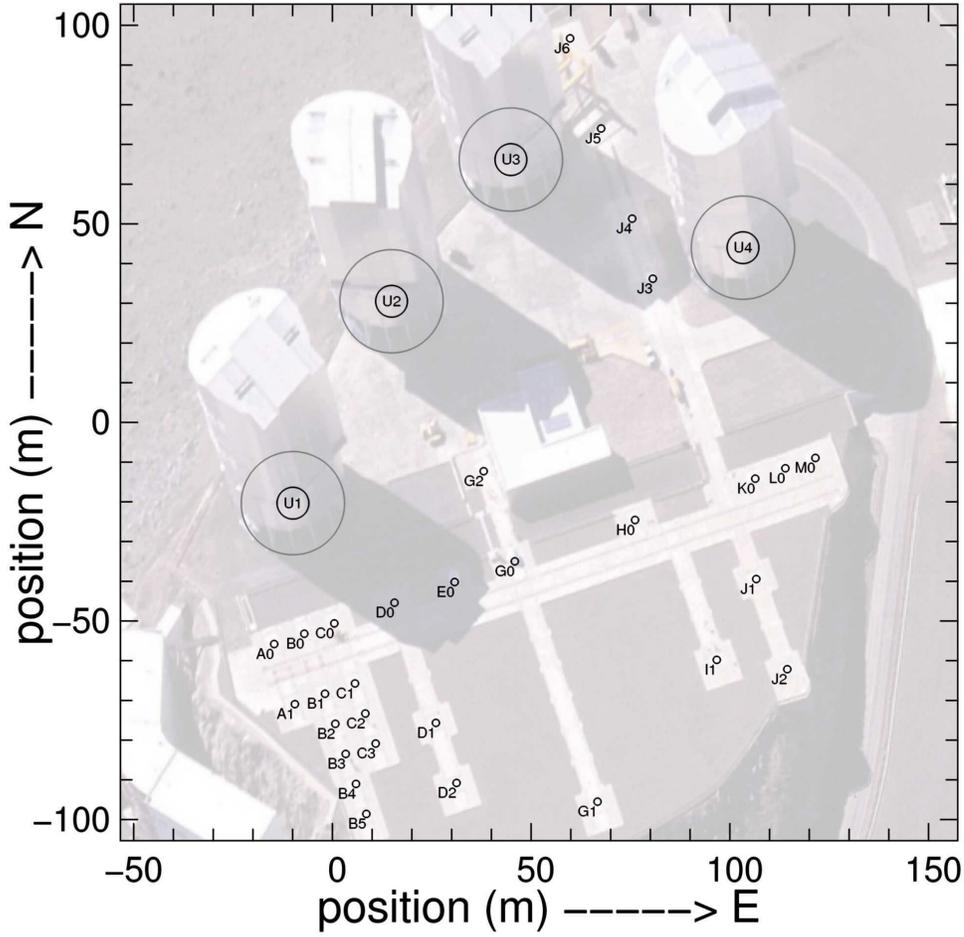}
  \caption[Paranal sketch]{
    \footnotesize{
      Aerial view of the Paranal mountain and the Very Large
      Telescopes (photo credit: Gerhard H\"udepohl) under-plotted
      on the VLTI stations. Small stations correspond to ATs and
      larger encircled ones correspond to UTs.
    }
  }
  \label{fig:Photo_VLTI}
\end{figure}

VLT (or \emph{Very Large Telescope}) is European project % leaded by
% ESO (\emph{European Southern Observatory}) with an annual budget of \EUR50
% millions. It is
mainly made of four 8\,m telescopes (Unit
Telescopes or UT) and a series of ``small'' 1.8\,m telescopes
(Auxiliary Telescopes or AT). Today, 4 ATs and 4 UTs
are operational with baselines ranging from 16\,m to 130\,m
(Fig.~\ref{fig:Photo_VLTI}).
Like the Keck telescope, the VLT can operate each 8\,m telescope
individually or combine up to three telescopes together using the
VLTI or \emph{Very Large Telescope Interferometer}. 3 instruments
are now operational at VLTI: 
\begin{description}
\item[VINCI:] test instrument,
\item[MIDI:] mid-infrared instrument using 2 telescopes
\item[AMBER:] 3 telescope instrument in the near infrared.
\end{description}
Today's VLTI is limited by vibrations,
which are under investigation and should be solved in the next few
years. The future of VLTI consists of a phase-referencing facility,
PRIMA, and a series of projects for second-generation instruments
(MATISSE, VSI, and GRAVITY).

\vspace{1cm}
\vspace{1cm}

% \subsection{Generic principle}

\subsection{Definitions}

In this section, I will use the following notations:
\begin{itemize}
\item $t$ is time,
\item $\tau$ is delay,
\item $\omega$ is the light pulsation and $\lambda$ is its wavelength:
  $\omega = 2 \pi C / \lambda$ 
\item $\overrightarrow{x}$ is the $(x,y,z)$ coordinates vector,
\item $\overrightarrow{p}$ is the $(x,y)$ coordinates vector in the
  pupil plane,
\item $\overrightarrow{s}$ is the $(x,y)$ coordinates vector in the
  detector plane,
\item $\overrightarrow{u}$ is the $(u,v)$ coordinates vector in the
  Fourier plane.
\end{itemize}

Other notations will be defined as they appear.

\paragraph*{Electromagnetic wave:}

An electromagnetic wave $(\overrightarrow{E}(\overrightarrow{x},t),
\overrightarrow{B}(\overrightarrow{x},t))$ is a particular case of an
electric and magnetic field that propagates in space with
time. The propagation of an electromagnetic wave is described by the
Maxwell equations, and using the Lorentz gauge in space, it can take
the following form:

\begin{eqnarray}
  \overrightarrow{E}(\overrightarrow{x},t) & = &
  \overrightarrow{E_0}(\overrightarrow{x}) e^{\imath \omega t}\\
  \overrightarrow{B}(\overrightarrow{x},t) & = &
  \overrightarrow{B_0}(\overrightarrow{x}) e^{\imath \omega t}
  \nonumber
  \label{eq:onde_electromagnetique}
\end{eqnarray}

\paragraph*{Light intensity:}

An optical or infrared detector is sensitive to the light
intensity $I(\overrightarrow{s})$; i.e. the time-average of the
squared modulus of the electromagnetic field at the measurement point:

\begin{equation}
  I(\overrightarrow{s}) =
  \left< \left\| \overrightarrow{E}(\overrightarrow{s},t) \right\|^2 \right>_t
\end{equation}

\begin{figure}[htbp]
  \centering
  \includegraphics[width=0.95\textwidth]{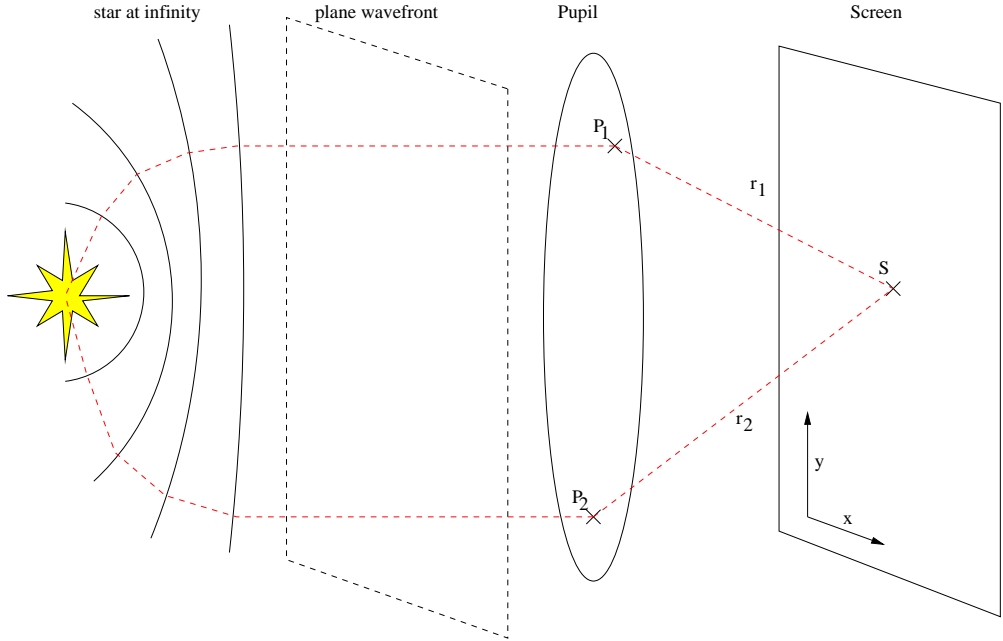}
  \caption[Light propagation of an electromagnetic wave]{
    \footnotesize{
      The light propagation of an electromagnetic wave from a star at
      infinity to the detector plane  (screen). Interferometry is
      interested in the study of the interaction between waves coming
      from two points of  the initial wavefront. Notations come from
      the text.
    }
  }
  \label{fig:propagation}
\end{figure}

In this equation, $\overrightarrow{s}$ are the spatial coordinates in
the detector plane (see Fig.~\ref{fig:propagation}), $t$ is the time,
and $\overrightarrow{E}$ is the electromagnetic field.

\vspace{1cm}
\vspace{1cm}

\subsection{What does an interferometer measure ?}

\paragraph*{Coherence between 2 waves:}

When observing an object with a given instrument, the light intensity
$I$ is the result of the superposition of many electromagnetic waves
coming from different points of the instrument's pupil plane (see
Fig.~\ref{fig:propagation}):

\begin{eqnarray}
  I(\overrightarrow{s}) & = &
  \left< \left\| \overrightarrow{E}(\overrightarrow{s},t) \right\|^2 \right>_t\\
  & = &
  \left< \left\| \sum_i \overrightarrow{E}(\overrightarrow{p_i},t - \tau_i)
    \right\|^2 \right>_t
  \nonumber
\end{eqnarray}

This expression applies when one is only interested in a 2-wave
interaction:

\begin{equation}
  I(\overrightarrow{s}) =  \left< \left\|
      \overrightarrow{E}(\overrightarrow{p_1},t - \tau_1) +
      \overrightarrow{E}(\overrightarrow{p_2},t - \tau_2) \right\|^2 \right>_t
\end{equation}

If one defines the mutual coherence function $\Gamma_{1,2}$ by:

\begin{equation}
  \Gamma_{1,2}(\tau) =  \left<
    \overrightarrow{E}(\overrightarrow{p_1},t - \tau)
    \overrightarrow{E}^{*}(\overrightarrow{p_2},t) \right>_t
  \label{eq:coherence_mutuelle}
\end{equation}

The light intensity can be expressed in a simple way:

\begin{equation}
  I(\overrightarrow{s}) = \Gamma_{1,1}(0) + \Gamma_{2,2}(0) +
  \Gamma_{1,2}(\tau_2 - \tau_1) + {\Gamma_{1,2}}^{*}(\tau_2 - \tau_1)
\end{equation}

Then one can define the complex coherence degree $\gamma_{1,2}(\tau)$
as:

\begin{equation}
  \gamma_{1,2}(\tau) =
  \frac{\Gamma_{1,2}(\tau)}{\Gamma_{1,1}(0) + \Gamma_{2,2}(0)}
\end{equation}

Light intensity becomes:

\begin{equation}
  I(\overrightarrow{s}) = \left[ I_1(\overrightarrow{s}) +
    I_2(\overrightarrow{s}) \right] \left[ 1 + \Re \left(
      \gamma_{1,2}(\tau) \right) \right]
\end{equation}

which, in the case of wave-fronts at infinity, as described in equation
\ref{eq:onde_electromagnetique}, and considering the on-screen
coordinate $x$ and the wavelength $\lambda = 2 \pi c / \omega$, the
light intensity becomes:

\begin{equation}
  I(\overrightarrow{s}) = \left[ I_1(\overrightarrow{s}) +
    I_2(\overrightarrow{s}) \right] \left[ 1 + \mu \cos \left( \frac{2 \pi
        x}{\lambda} + \phi \right) \right]
  \label{eq:interferogramme}
\end{equation}

$\mu$ being the modulus of $\gamma_{1,2}(0)$ and $\phi$ its phase.

\vspace{1cm}
\begin{figure}[htbp]
  \centering
  \begin{tabular}{cc}
    \includegraphics[height=0.4\textwidth]{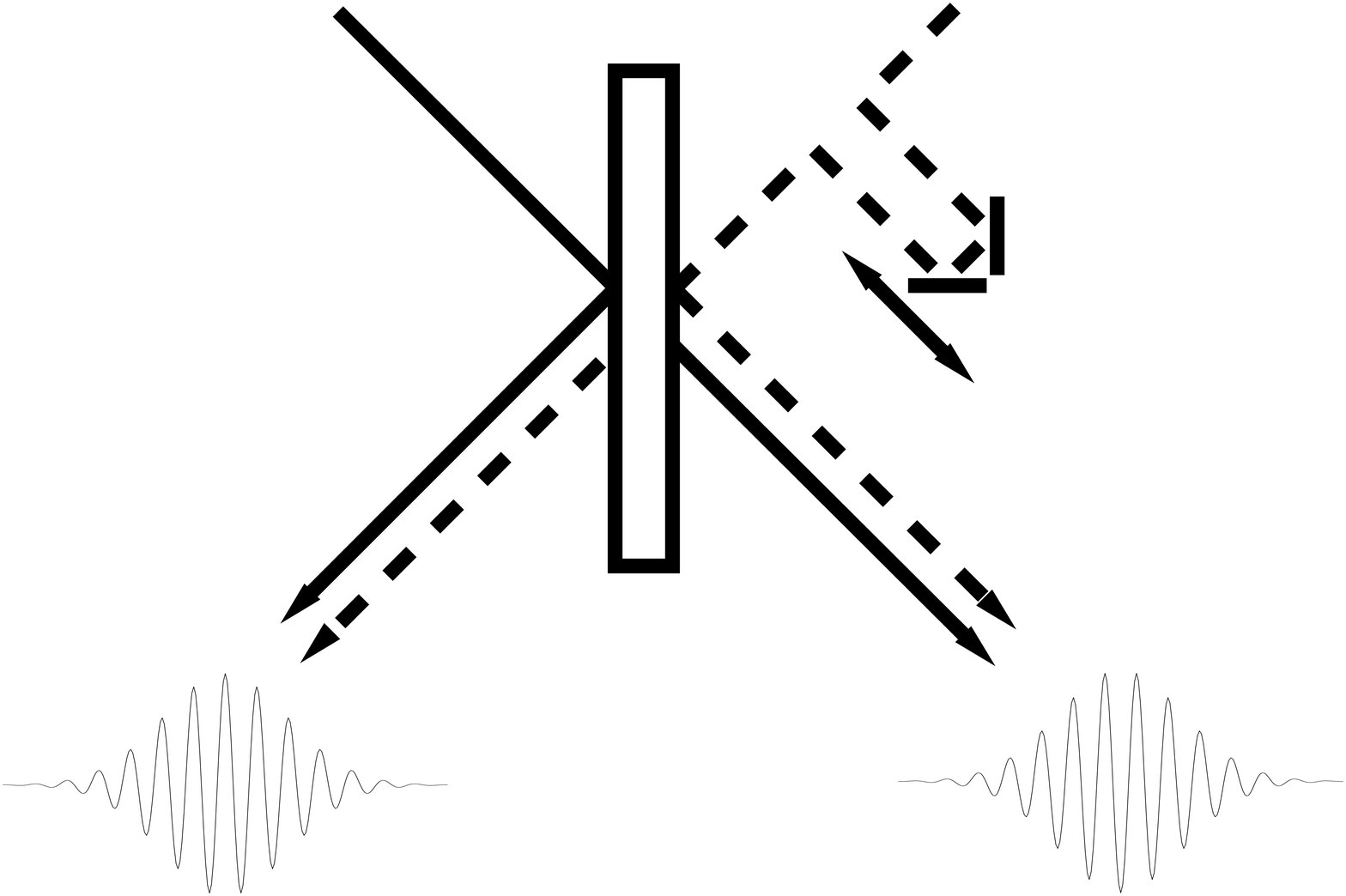}&
    \includegraphics[height=0.4\textwidth]{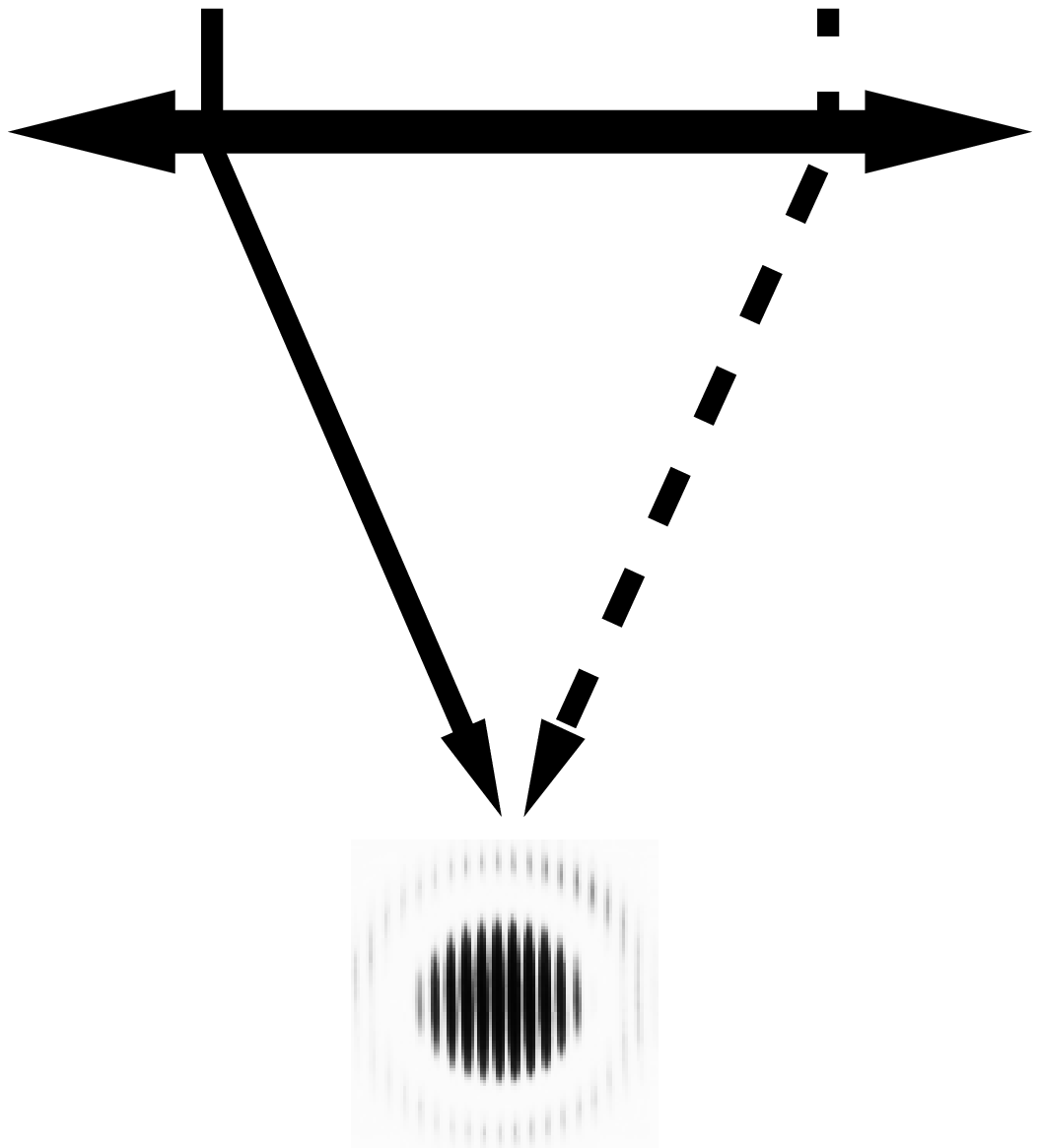}
  \end{tabular}
  \caption[Coaxial / Multiaxial]{
    \footnotesize{
      Coaxial vs multiaxial beam recombination. In co-axial
      recombination, the fringes are scanned by a temporal modulation,
      whereas for multiaxial, the different pixels of the detector
      scans different OPDs. The nature of the recorded signal is then
      different and different biases affect them.
    }
  }
  \label{fig:coaxMultiax}
\end{figure}
\vspace{1cm}

This co-sinusoidal modulation of the light is called an interferogram
and is what one can measure using an interferometer ($x$ is then
modulated spatially - multiaxial instrument - or temporally - coaxial
instrument - see Fig.~\ref{fig:coaxMultiax}).

\subsubsection*{The Van-Cittert / Zernike theorem:}
\label{sbsbsect:Zernike}

The Van-Cittert / Zernike theorem describes the relation between what
we call the ``complex visibility'' (i.e. $\mu_{1,2} =
\gamma_{1,2}(0)$) of an object and its brightness distribution
$o(\overrightarrow{\alpha})$ on the plane of the sky. One can find its
demonstration in various books, such as \citet{1985_goodman}.

\paragraph*{Theorem:}
\emph{For a non-coherent and almost monochromatic extended source, the
  complex visibility is the normalised Fourier transform (hereafter
  FT) of the brightness distribution of the source.}

\begin{figure}[htbp]
  \centering
  \begin{tabular}{ccc}
    % \multicolumn{2}{c}{\includegraphics[width=0.48\textwidth]{hip4849_ima.ps}}\\
    % \includegraphics[width=0.48\textwidth]{hip4849_re.ps}&
    % \includegraphics[width=0.48\textwidth]{hip4849_im.ps}\\
    \includegraphics[width=0.32\textwidth]{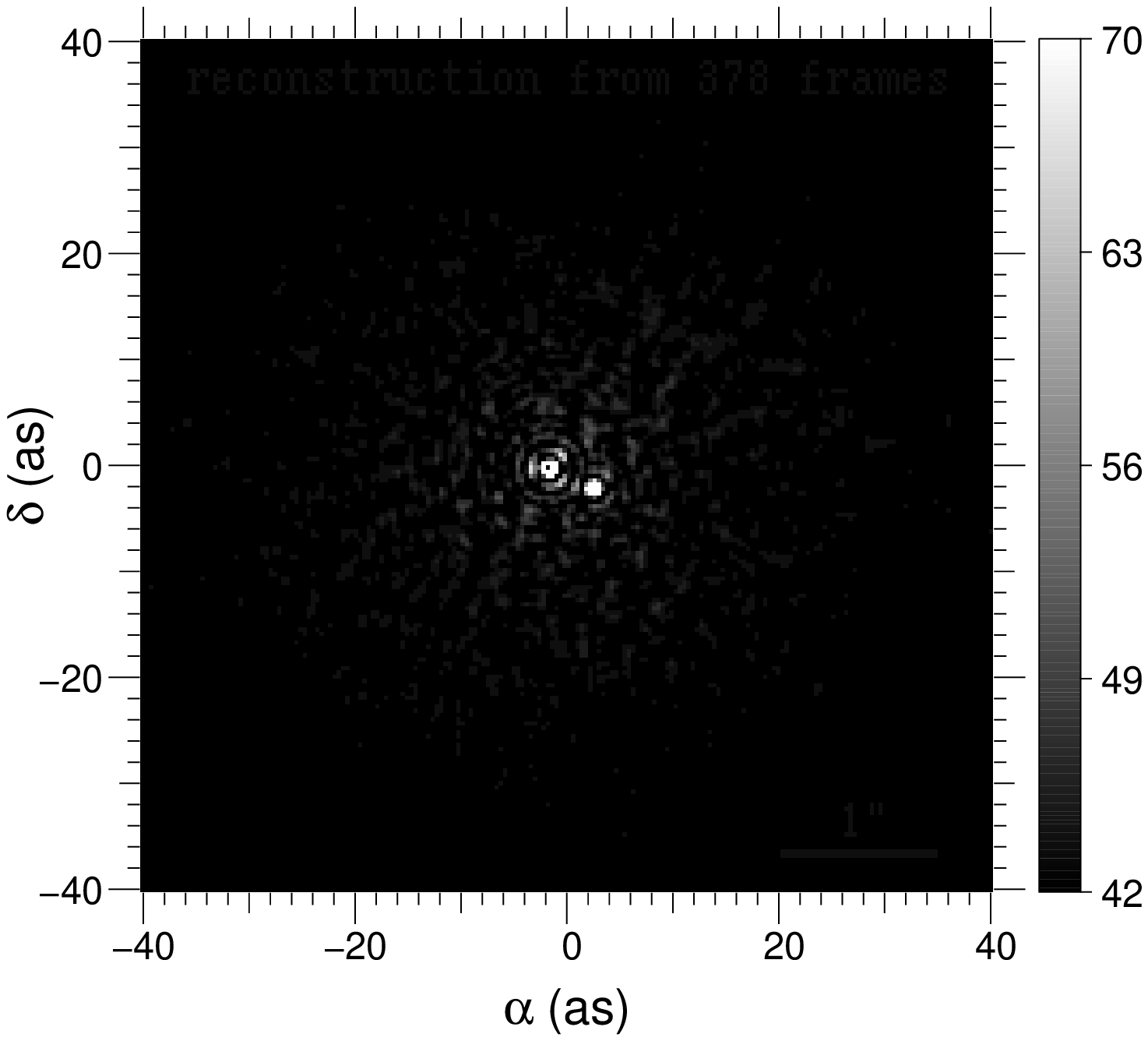}&
    \includegraphics[width=0.32\textwidth]{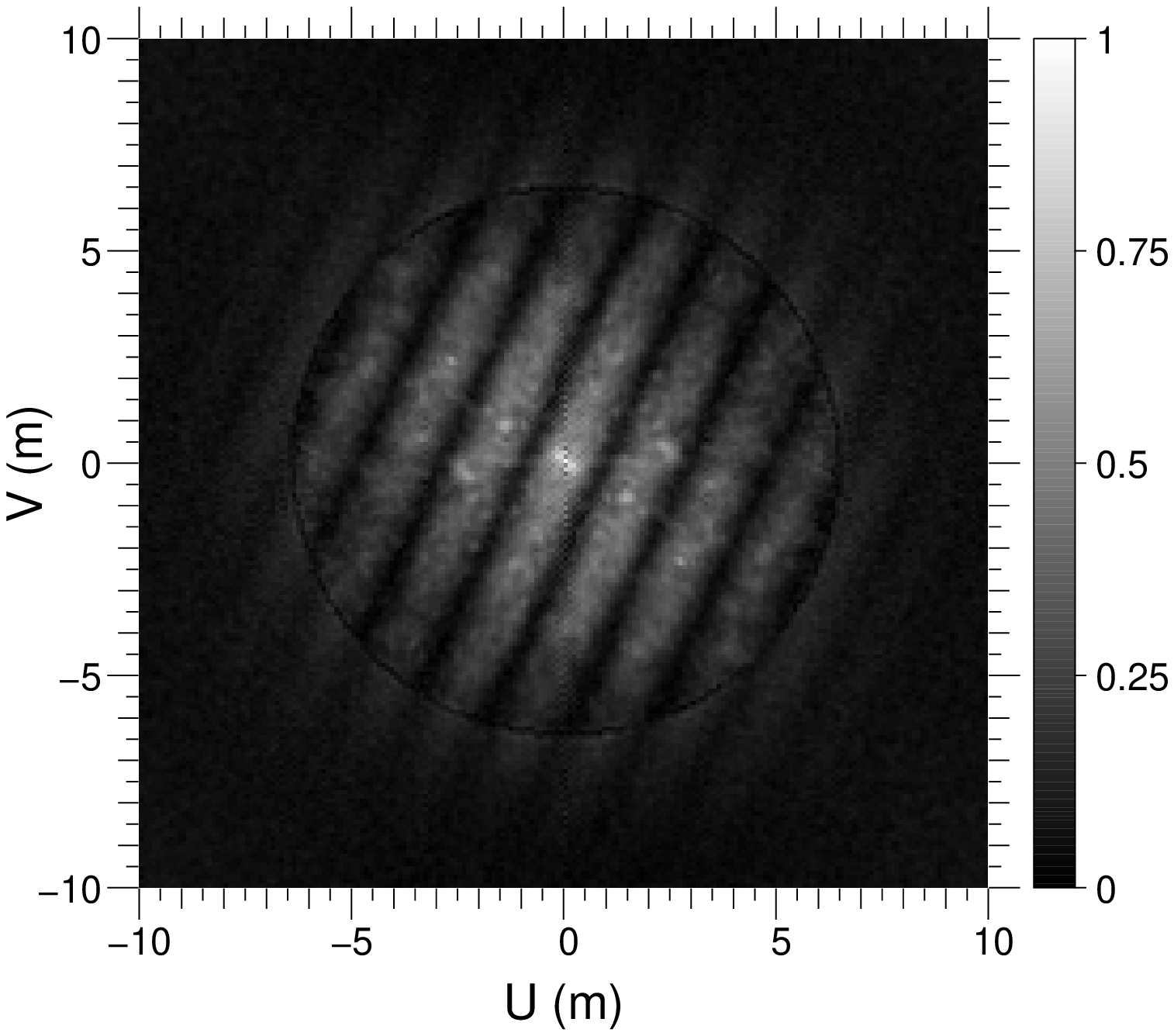}&
    \includegraphics[width=0.32\textwidth]{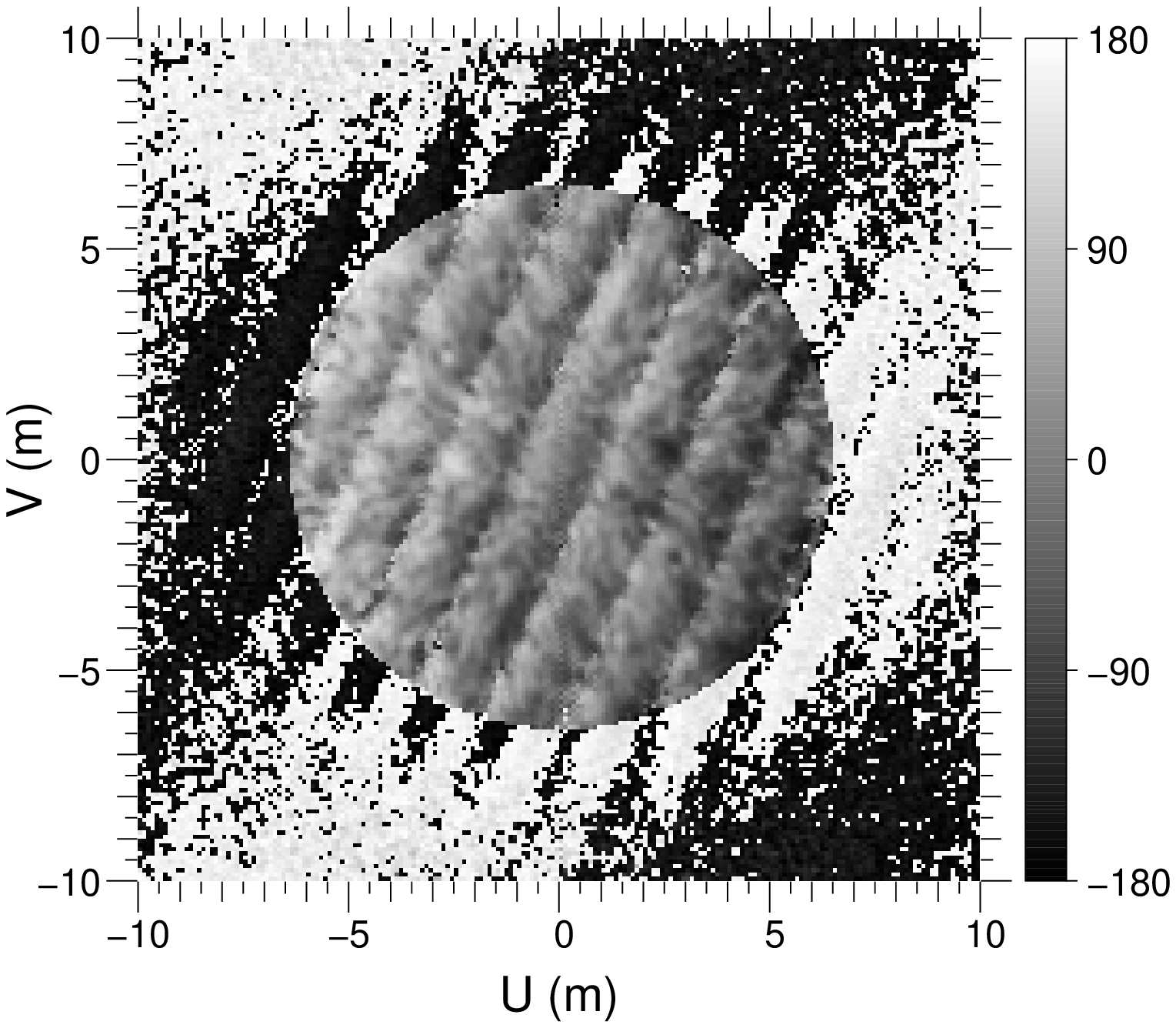}
  \end{tabular}
  \caption[Super-synthesis illustration]{
    \footnotesize{
      Example \citep[taken
      from][% \footnote{\url{http://www.mpifr-bonn.mpg.de/old_mpifr/div/ir-interferometry/}}
      ]{2006A&A...448..703B}
      of a diffraction-limited image of a binary star (HIP 4849,
      observed at the Special Astronomical Observatory, Zelentchouk,
      left). The complex visibility of this image (middle and right)
      gives typical waves until the cut-off frequency of 6\,m
      (features at larger frequencies are produced by the image
      reconstruction algorithm used here). 
    }
  }
  \label{fig:UV_coverage}
\end{figure}

By definition, the visibility $\mu$ is a complex number, whose modulus is
between 0 and 1 (see Fig.~\ref{fig:UV_coverage} for an example in the
case of a mono-pupil instrument). It is by definition:

\begin{equation}
  \mu(\overrightarrow{u}) =
  \frac{\widetilde{o}(\overrightarrow{u})} {\iint
    o(\overrightarrow{\alpha})}
\end{equation}

where\  $\ \ \widetilde{ }\ \ $\  represents the FT. By definition,
when one measures $\|\mu\| < 1$, the object observed is
\emph{resolved} by the instrument. We can see here that interferometry
does not provide direct access to the image of an object, as single-dish
telescopes
do. It is sensitive directly to the FT of the

brightness distribution of the object.

\subsection{Practical considerations}

Optical long-baseline interferometry has several intrinsic difficulties:
\begin{enumerate}
\item the intrinsic complexity to measure the data,
\item the apparent complexity for understanding the interferometric
  measures,
\item the very sparse sampling of data,
\item the relative bad sensitivity compared to
  single-dish experiments.
\end{enumerate}

\subsubsection{Measurement complexity}
\begin{figure}[htbp]
  \centering
  \includegraphics[width=0.95\textwidth, angle=0]{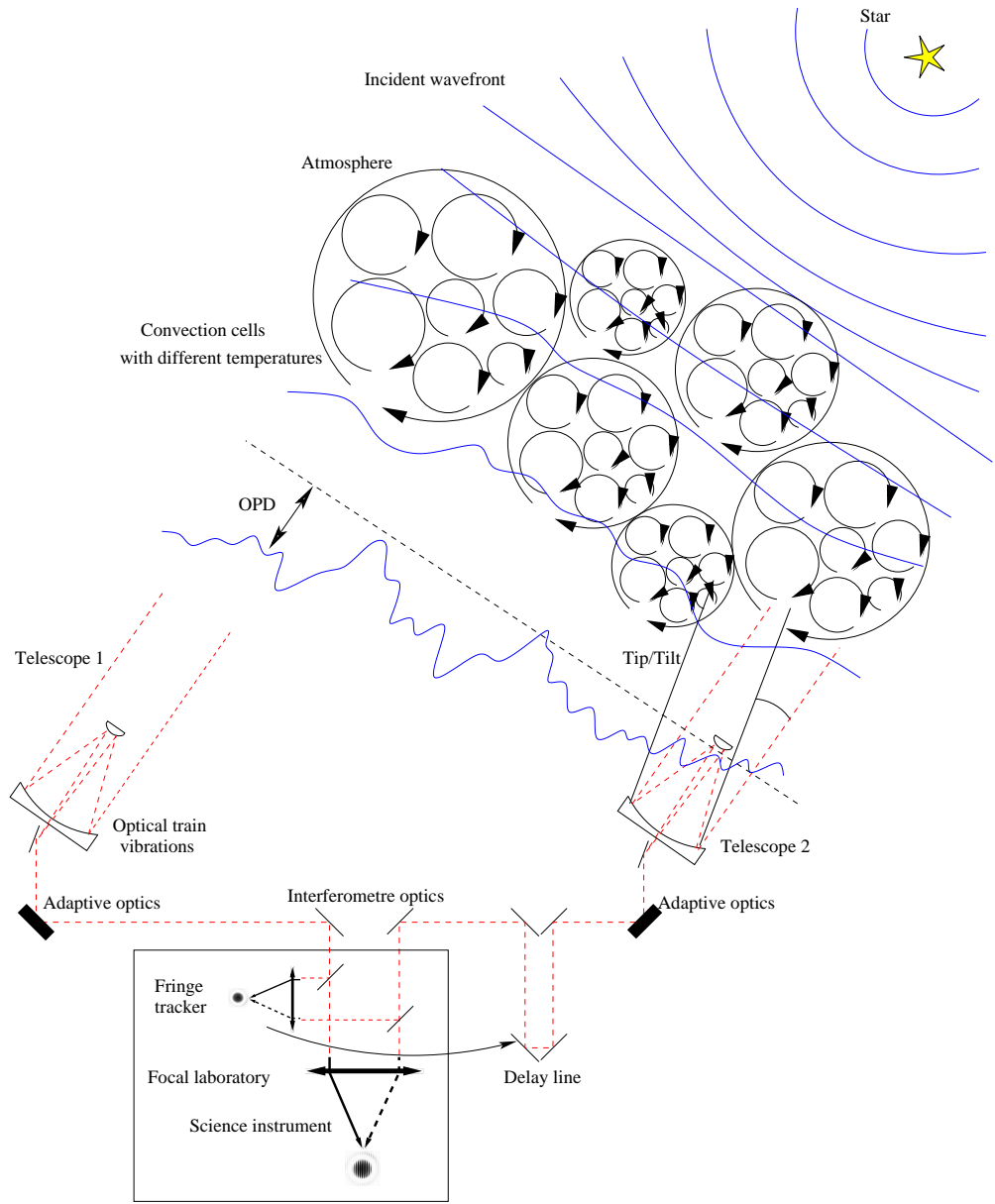}
  \caption[Interferometre general scheme]{
    \footnotesize{
      This figure shows the main effects affecting an interferometric
      measurement: the atmospheric OPD being the most
      dominant by far. It also shows today's solutions to correct
      these effects: adaptive optics (or a tip/tilt correction for
      small apertures) for telescope effects, and fringe tracker
      and delay lines for interferometer effects.
    }
  }
  \label{fig:interfScheme}
\end{figure}

The first problem can be divided into two main problems:
\begin{itemize}
\item The atmospheric effect on ground-based observatories produces
  very fast and varying phase variations. In the case of single-dish
  telescopes, this effect typically reduces the angular resolution to
  the one of a telescope of the Fried diameter $r_0$
  \citep{1965_Fried}. Today, the advent of adaptive optics allows one
  to reach much  higher angular resolutions at the cost of monitoring the
  atmosphere effects very fast and limiting the accessible magnitude. In the
  case of interferometry, long exposure times are not possible due to
  this atmospheric phenomenon. Therefore, just like with adaptive
  optics, a fast feedback loop must operate to correct the fringes
  motion. This is the only way of getting longer exposure times with
  the science instrument (see Fig.~\ref{fig:interfScheme}).
\item The use of many subsystems in an interferometer (telescopes,
  delay lines, image sensors, fringe sensors, adaptive optics, focal
  instruments, etc.) can also be an issue. The more complex the interferometer
  gets, the higher the probability of failure. System
  optimisation is then a key  point to manage these complex systems.
\end{itemize}

\vspace{1cm}
\vspace{1cm}

\subsubsection{Interferometry understanding} 
The second problem  is about to be solved with the development of general
astrophysics instruments (AMBER is an example), and the
preparation and interpretation of observations can be done with easy-to-use tools
(developed by ESO\footnote{European Southern Observatory}, JMMC
\footnote{Jean-Marie   Mariotti Centre}, and MSC \footnote{Michelson
  Science Centre}). For the general user, continuous trainings and
summer schools are intended to help with the data interpretation.

\subsubsection{Sparse sampling} 
The third problem is also a 2-part problem:
\begin{itemize}
\item The limited number of telescopes (currently 3 on the VLTI, 4 with
  the second generation, 6 maximum for CHARA) generates a very sparse
  sampling of the (u,v) plane.
\item The atmosphere makes it very difficult to access the object's
  phase and prevents a direct measurement of it. Several methods
  exist to partially retrieve the phase:
\begin{itemize}
\item  the phase closure (starting
  from 3 telescopes),
\item  the differential phase (using a spectrograph),
\item  the phase reference.
\end{itemize}
Today, the phase closure is the most widely
  used since it is easiest to calibrate.
\end{itemize}
The strategy to observe and interpret the data is then very different
from what can be done with imaging facilities (such as single-dish
telescopes, at a lower spatial resolution).

\subsubsection{Low sensitivity} 
The last problem is also about to disappear with the advent of new
generation interferometers like Keck-I or VLTI, combining long
baselines and very large apertures.

\section{(u,v) plane properties}
\label{sect(uv)}

As was seen in the previous section, an interferometer is
sensitive to the FT of an object's brightness distribution. The
question of Fourier plane (also called (u,v) plane) sampling is then
crucial to know what part of the object's information the observer
really observed. This section explains how to get a sufficiently good
(u,v) coverage for an observation using a given interferometer.

\subsection{Super-synthesis}

The idea behind this word is to use Earth
rotation to get a larger (u,v) sampling. Since a stellar interferometer
baseline is fixed on the ground, its projection on the sky plane
changes as the Earth rotates. This baseline projection depends only on
the hour angle $h$ (i.e. $h =$ LST - R.A.), the baseline
coordinates $(X, Y, Z)$ and the declination of the object
$\delta$. A change of coordinates links the baseline
position and its projection on the plane of the sky:

\begin{eqnarray}
  \left(
    \begin{array}{c}
      u\\
      v\\
      w\\
    \end{array}
  \right)
  & = &
  \frac{1}{\lambda} 
  \left(
    \begin{array}{ccc}
      sin(h) & cos(h) & 0 \\
      -sin(\delta) cos(h) & sin(\delta) cos(h) & cos(\delta) \\
      cos(\delta) cos(h) & -cos(\delta) sin(h) & sin(\delta)
    \end{array}
  \right)
  \left(
    \begin{array}{c}
      X\\
      Y\\
      Z\\
    \end{array}
  \right) 
\end{eqnarray}

\begin{figure}[htbp]
  \centering
  \begin{tabular}{cc}
    \includegraphics[width=0.48\textwidth]{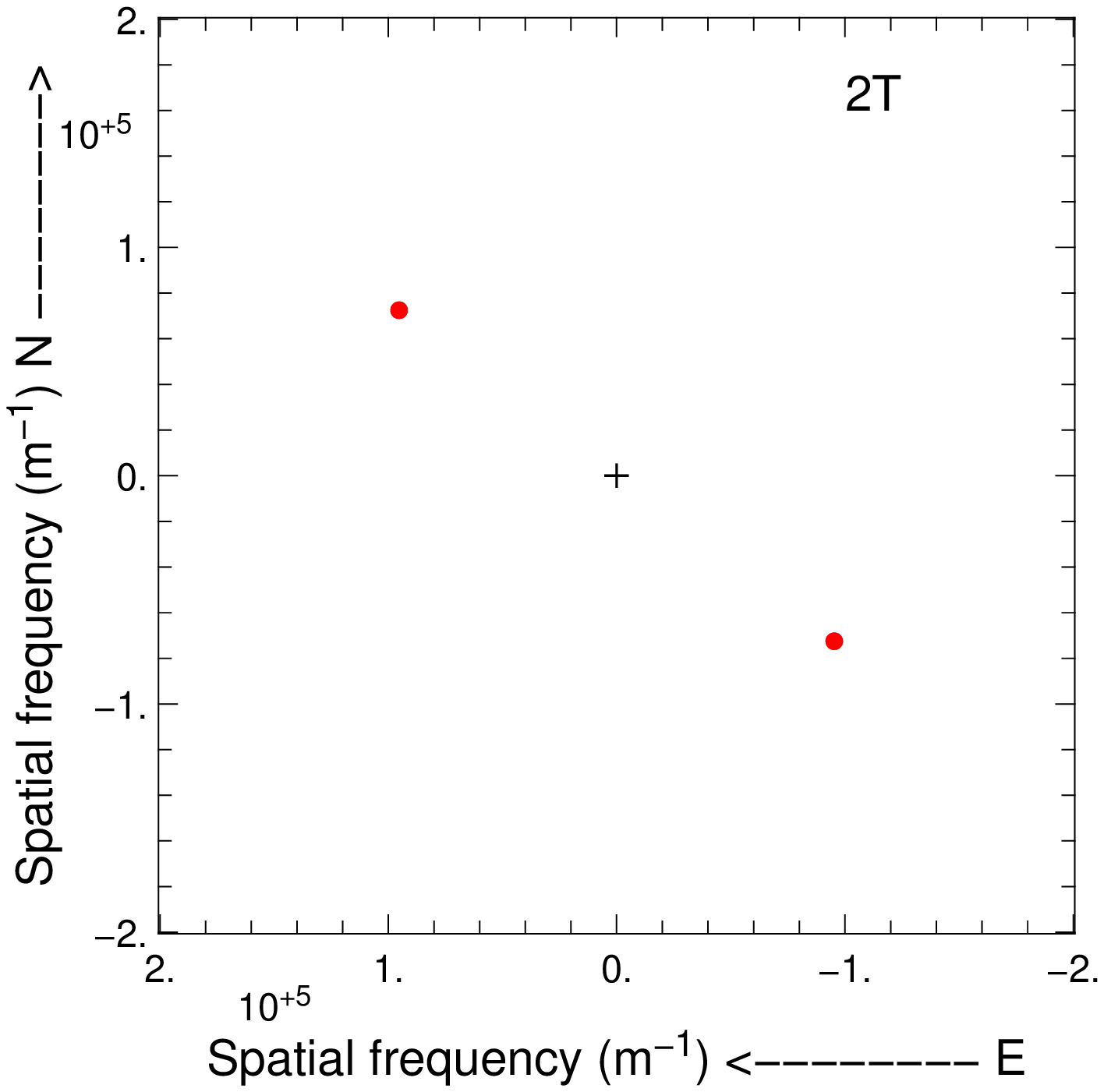}&
    \includegraphics[width=0.48\textwidth]{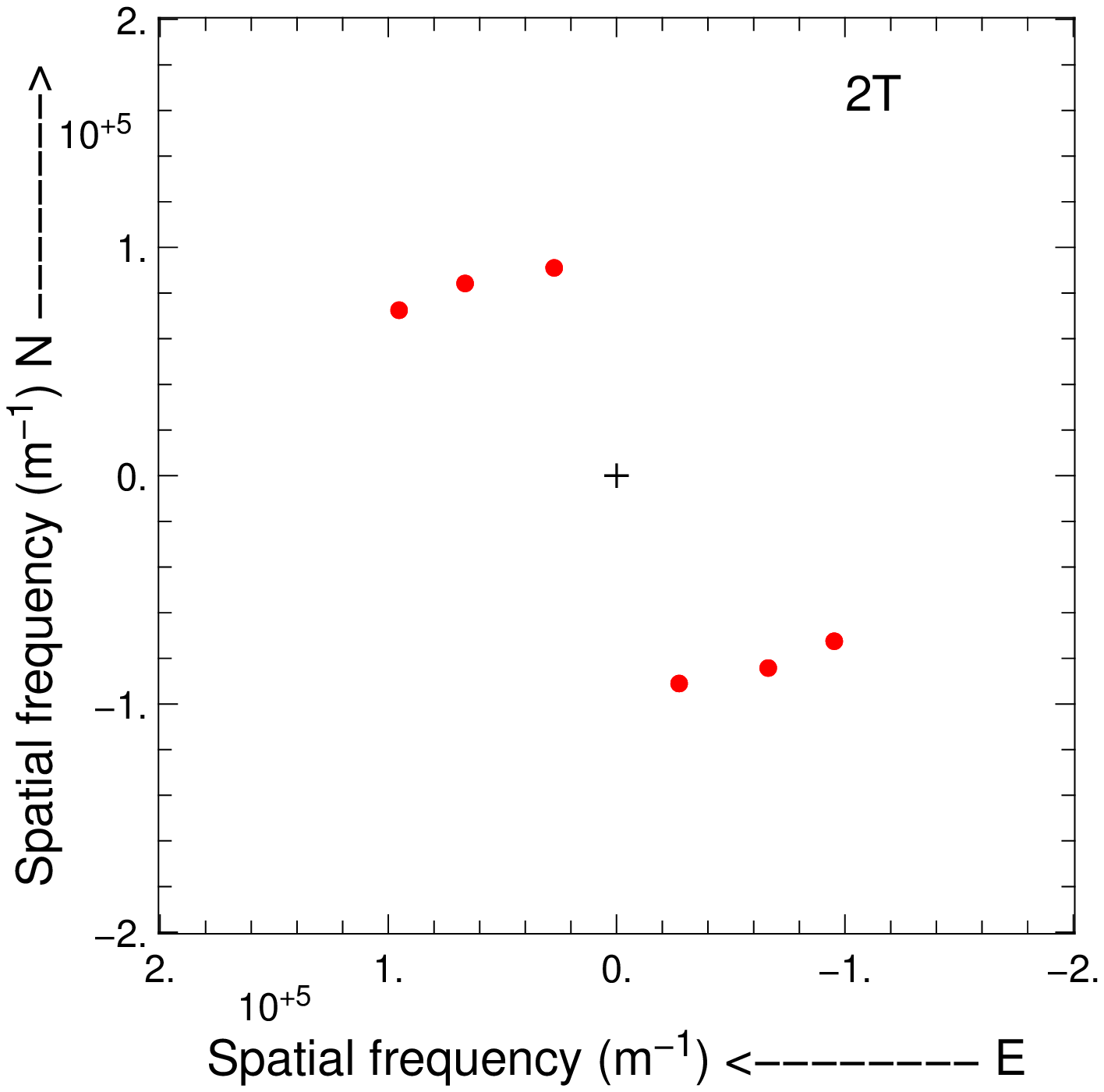}\\
    \includegraphics[width=0.48\textwidth]{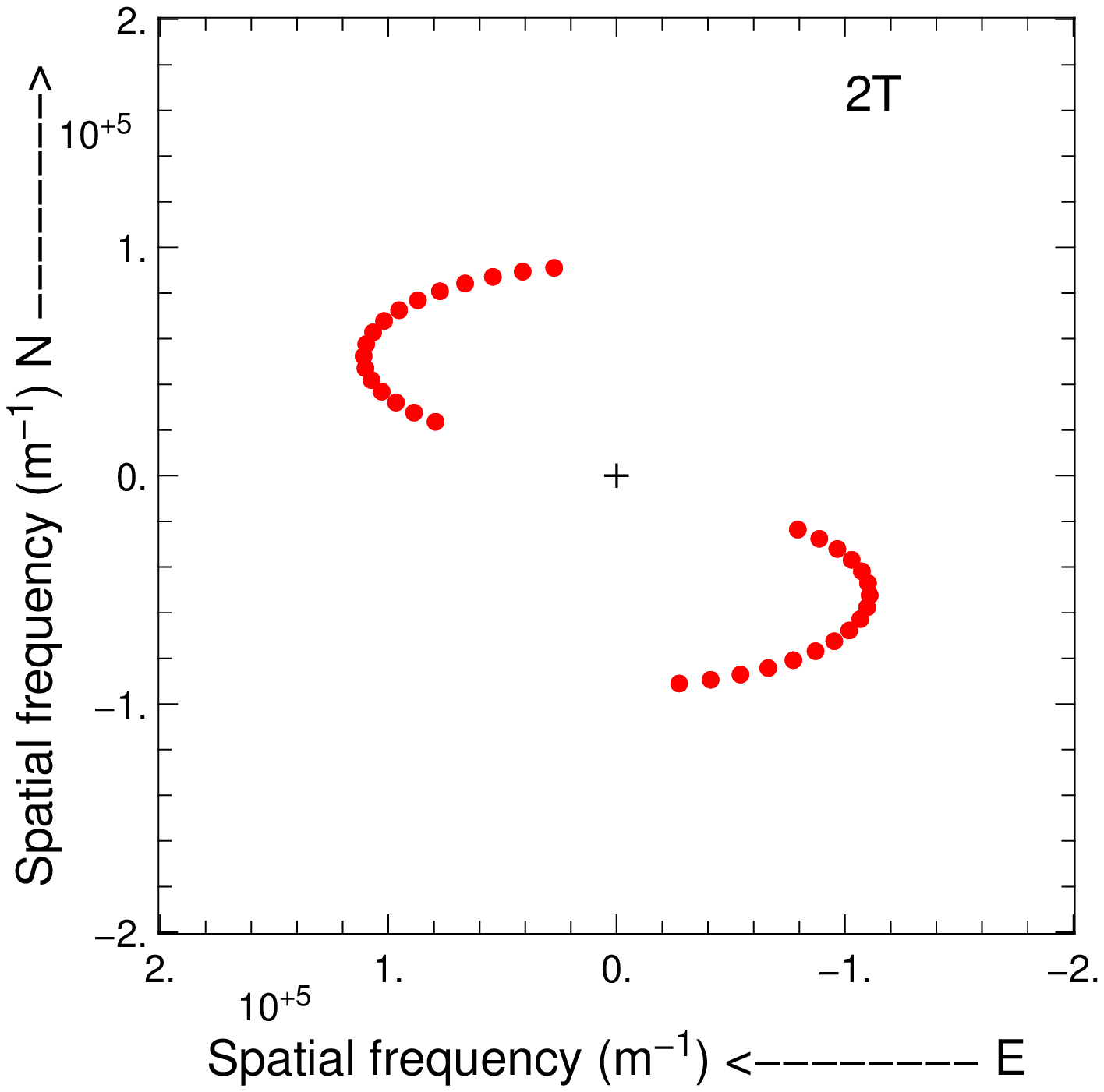}&
    \includegraphics[width=0.48\textwidth]{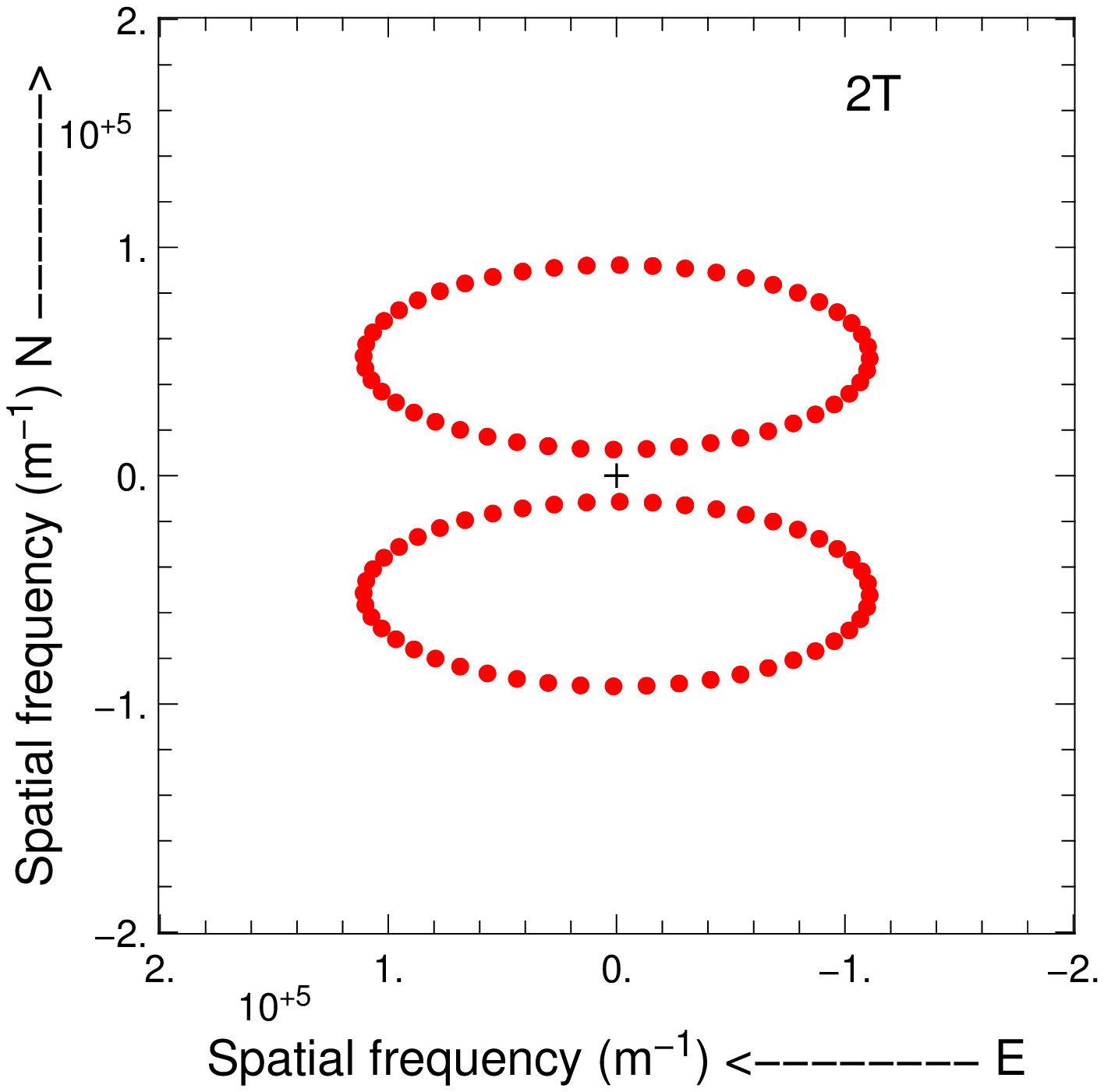}
  \end{tabular}
  \caption[Super-synthesis illustration]{
    \footnotesize{
      Starting from a snapshot with 2 telescopes (UT1-UT4, top left), and
      increasing both the time sampling (1.5h top-right and 1/2h
      bottom-left) and the total integration time (1/2 night top-right
      and 1 night bottom-left), one can see that the (u,v) path is an arc of
      an ellipse (bottom-right).
    }
  }
  \label{fig:2T_UV_coverage}
\end{figure}

The (u,v) coordinates for a given object ($\delta$ fixed)
depend only on a linear expression of $\cos(h)$ and $\sin(h)$. These
coordinates lies on an ellipse, as seen in
Fig.~\ref{fig:2T_UV_coverage}. The time spent on the object will determine the
increase in (u,v) coverage. However, the relation between the (u,v)
filling and this time span cannot be intuitively estimated. Therefore,
one has to check this for each studied case.

\vspace{1cm}
\vspace{1cm}

\subsection{Number of telescopes}

The relation between the number of telescopes combined and the
number of baselines is the following
\citep[See][]{2003RPPh...66..789M}:

\begin{equation}
  N_{b} = \frac{N_{t}\left(N_t-1\right)}{2}
  \label{eq:bases}
\end{equation}

The quantity measured is the amplitude and phase of the visibility
function at the given baselines. Therefore, the total number of
measurable information is:

\begin{equation}
  N_{m}^{tot} = N_{t}\left(N_t-1\right)
  \label{eq:mestot}
\end{equation}

However, in optical long baseline stellar interferometry, things
are not that simple. When using one spectral bin (for example,
with a large band instrument) the phase information on each baseline
is completely lost due to the atmosphere random phase noise
insertion. However, some phase information can be retrieved with the
phase closure when using 3 or more telescopes. The number of
phase closures measurable is:

\begin{equation}
  N_{c} = \frac{\left(N_{t}-1\right)\left(N_t-2\right)}{2}
  \label{eq:clos}
\end{equation}

One normally has access to
$\frac{N_{t}\left(N_t-1\right)}{2}$ amplitudes and
$\frac{\left(N_{t}-1\right)\left(N_t-2\right)}{2}$ phases in
interferometry. This gives the actual measurable number of
information:

\begin{equation}
  N_{m}^{act} = \left(N_{t}-1\right)^2
  \label{eq:mesTotClos}
\end{equation}

\begin{figure}[htbp]
  \centering
  \includegraphics[width=0.7\textwidth]{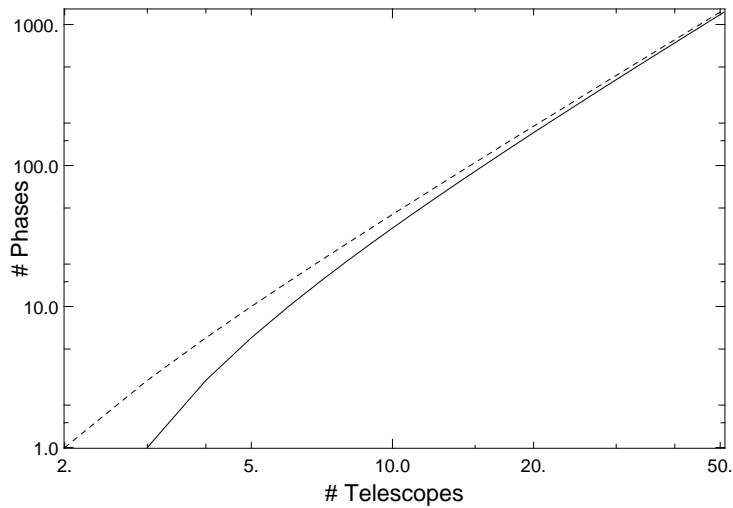}
  \caption[Increase of information by adding telescopes]{
    \footnotesize{
      One gain on 2 sides when adding telescopes to an optical long
      baseline stellar interferometer. The first is the squared increase
      of available baselines (dashed line).
      The second advantage is the phase closure information
      (full line) that grows faster for a few number of telescopes
      and tends to the maximum number of measurable phases (one per
      baseline, dashed line) for a high number of telescopes.
    }
  }
  \label{fig:basesClotures}
\end{figure}

The comparison between eq.\ref{eq:bases} and \ref{eq:clos} (see
Fig.~\ref{fig:basesClotures}) and also between eq.\ref{eq:mestot} and
\ref{eq:mesTotClos} suggest that the more
telescopes you have, the better you can measure
information at once.

\vspace{1cm}
\vspace{1cm}

% \begin{figure}[htbp]
%   \centering
%   \begin{tabular}{cc}
%     \includegraphics[width=0.48\textwidth]{UV_Coverage_2T_Gamma_Vel_1.05-1.05_U1U4_5_5_0.5.ps}&
%     \includegraphics[width=0.48\textwidth]{VLTI_stations_2T_Gamma_Vel_1.05-1.05_U1U4.ps}\\
% %     \includegraphics[width=0.48\textwidth]{UV_Coverage_3T_Gamma_Vel_1.05-1.05_U1U4U3_5_5_0.5.ps}&
% %     \includegraphics[width=0.48\textwidth]{VLTI_stations_3T_Gamma_Vel_1.05-1.05_U1U4U3.ps}\\
%     \includegraphics[width=0.48\textwidth]{UV_Coverage_4T_Gamma_Vel_1.05-1.05_U1U4U3U2_5_5_0.5.ps}&
%     \includegraphics[width=0.48\textwidth]{VLTI_stations_4T_Gamma_Vel_1.05-1.05_U1U4U3U2.ps}\\
%     \includegraphics[width=0.48\textwidth]{UV_Coverage_8T_Gamma_Vel_1.05-1.05_U1U4U3U2A1L0G1J6_5_5_0.5.ps}&
%     \includegraphics[width=0.48\textwidth]{VLTI_stations_8T_Gamma_Vel_1.05-1.05_U1U4U3U2A1M0G1J6.ps}
%   \end{tabular}
%   \caption[Increase of information by adding telescopes]{
%     \footnotesize{
%       Illustration of the interest in combining more and more
%       telescopes in stellar interferometry: the number of available
%       baselines increases much faster than the number of telescopes
%       added to the array (here from top to bottom, the telescopes
%       number is multiplied by 2 between each row). Left is the
%       simulation of a snapshot in the (u,v) plane and Right is the
%       corresponding baselines. Big dots are UTs and small ones are
%       ATs.
%     }
%   }
%   \label{fig:coherence_temporelle}
% \end{figure}

\subsection{Spectral coverage}

\vspace{1cm}
\begin{figure}[htbp]
  \centering
  \begin{tabular}{cc}
    \includegraphics[width=0.45\textwidth]{UV_Coverage_2T_Gamma_Vel_1.05-1.05_U1U4_5_5_0.5.ps}&
    \includegraphics[width=0.45\textwidth]{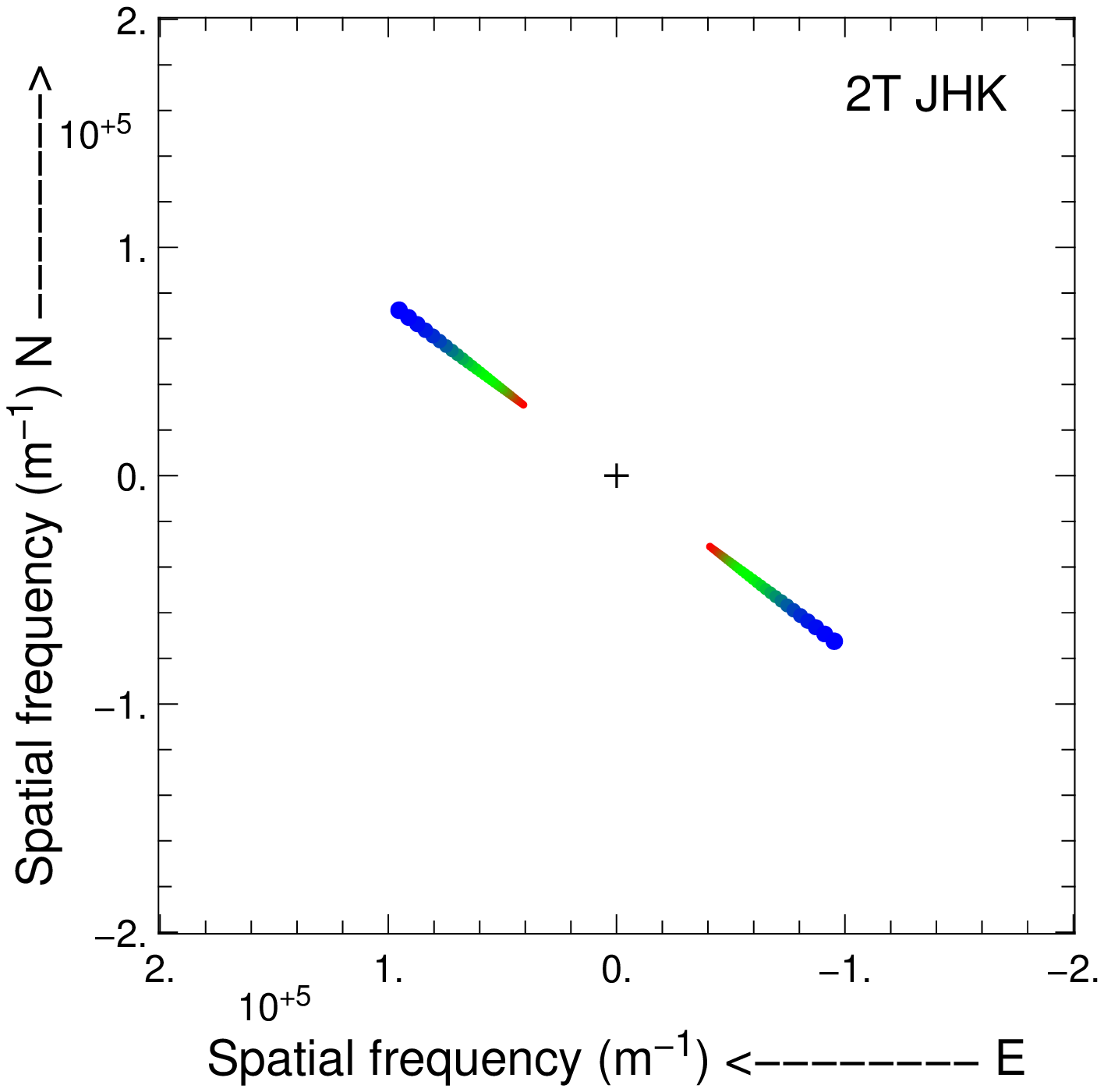}\\
    \includegraphics[width=0.45\textwidth]{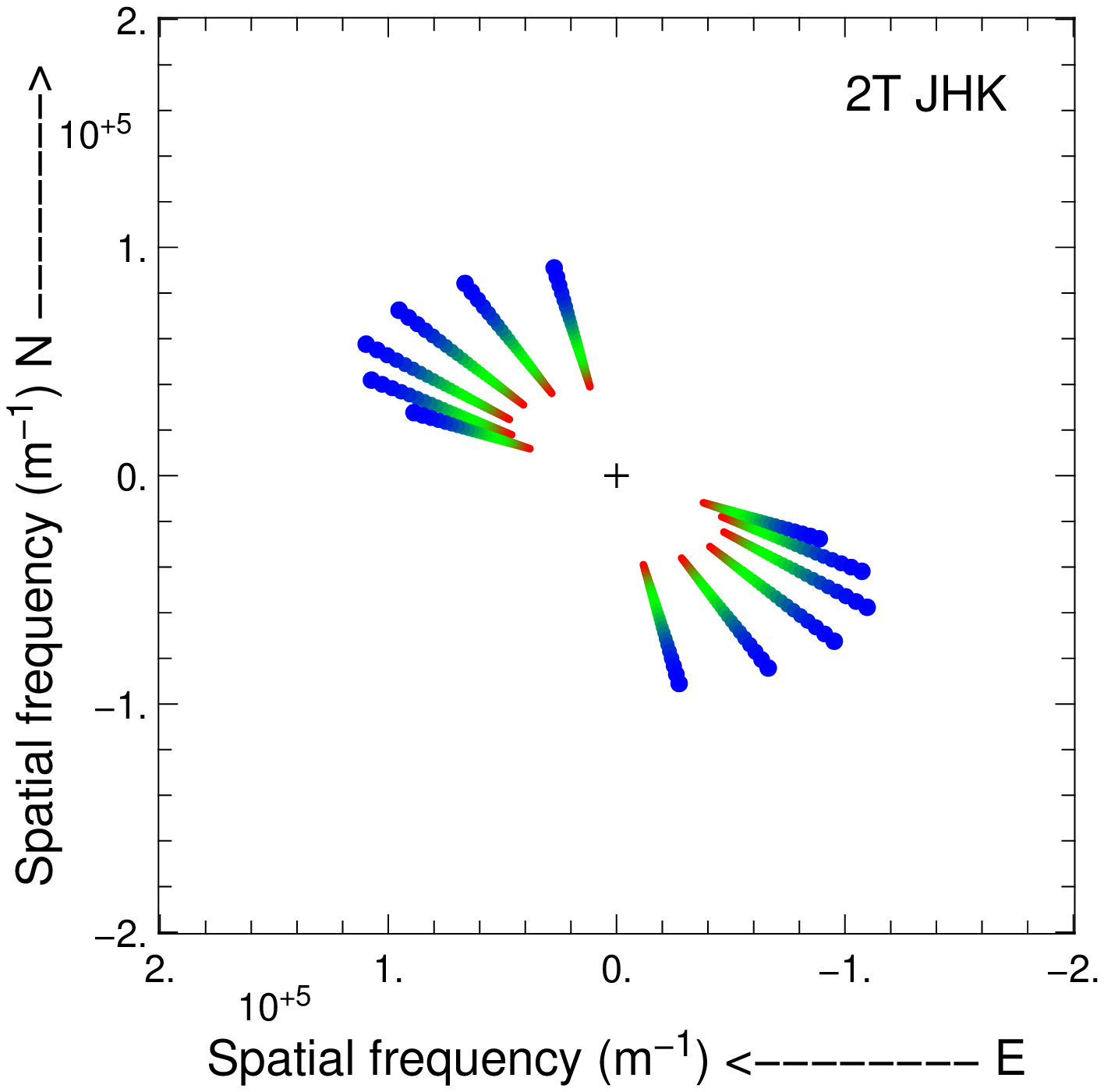}&
    \includegraphics[width=0.45\textwidth]{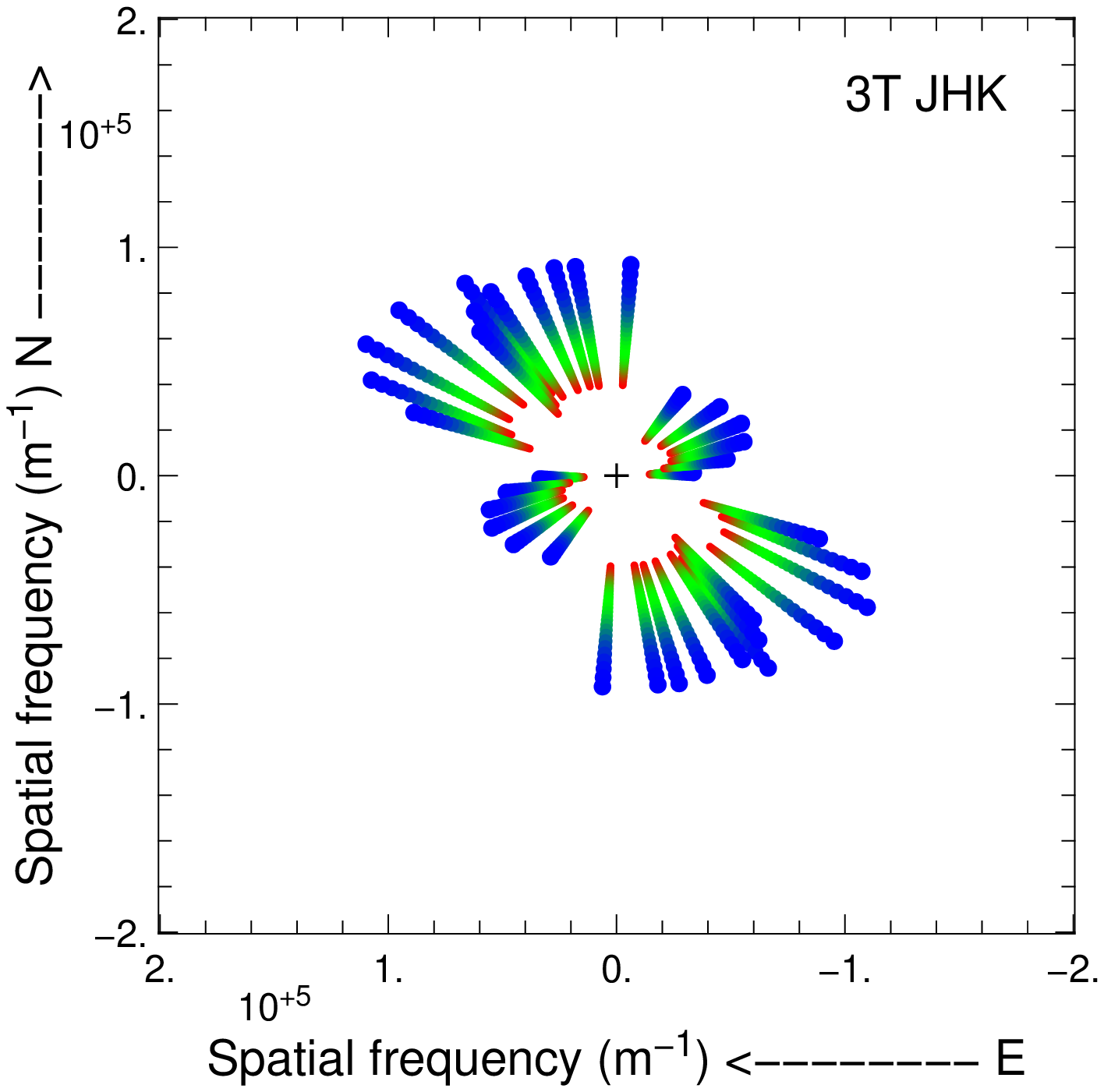}\\
    \includegraphics[width=0.45\textwidth]{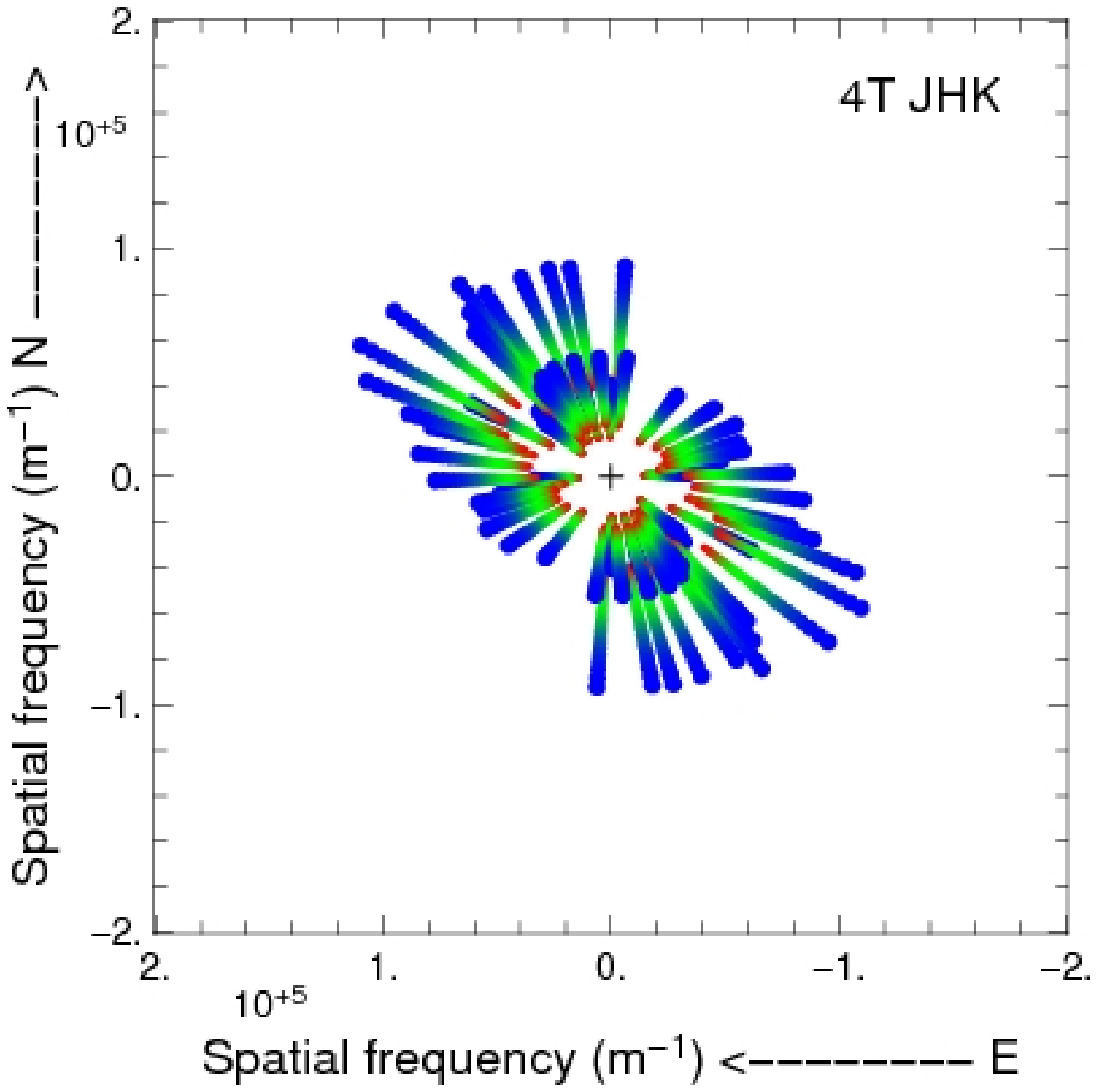}&
    \includegraphics[width=0.45\textwidth]{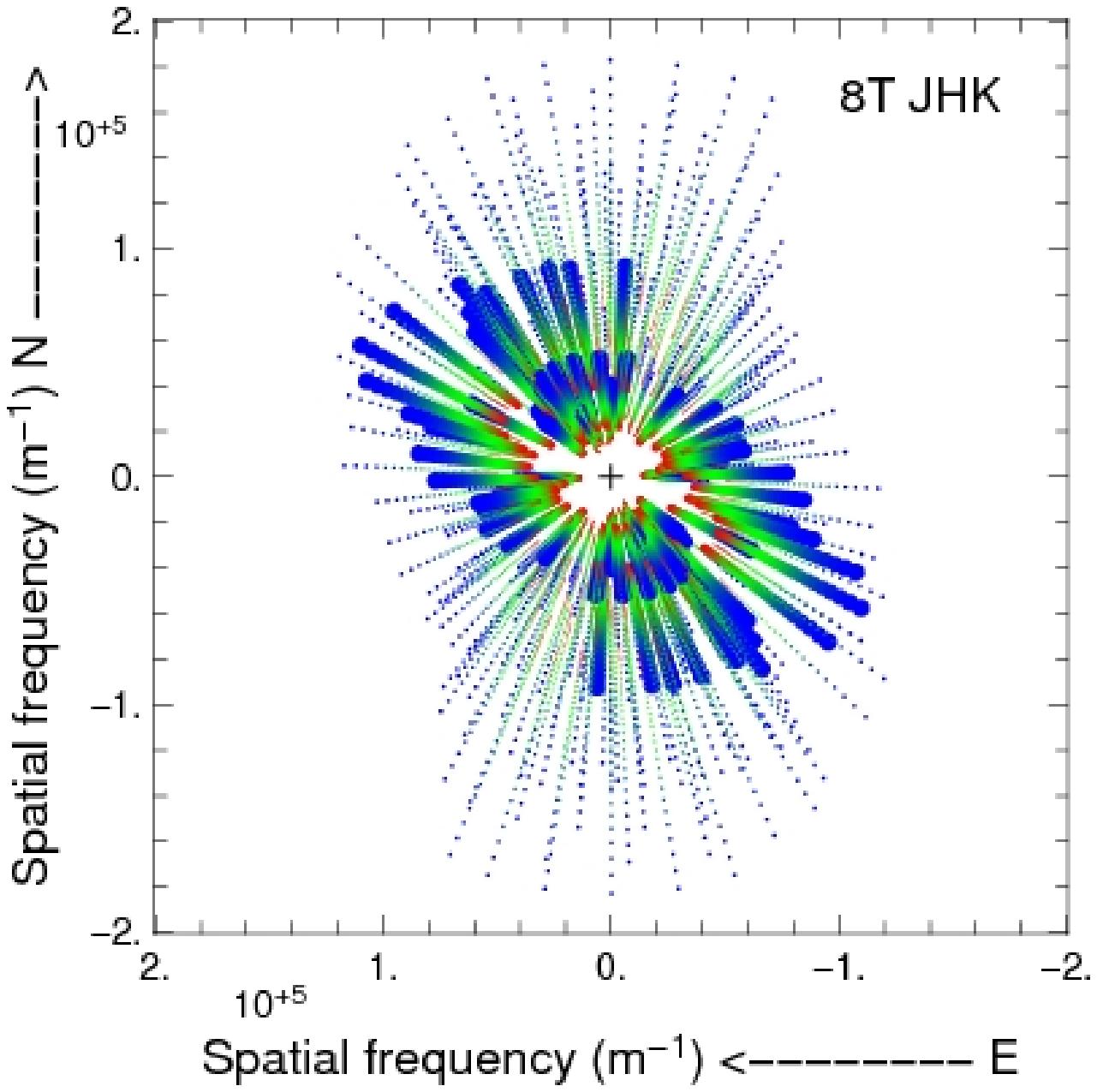}

  \end{tabular}
  \caption[Speckle binary]{
    \footnotesize{
      \emph{\bf Top-left:} (u,v) coverage of a single snapshot with a
      single spectral bin instrument (ex: VINCI).\emph{\bf Top-right:}
      The same but using spectroscopic capability (ex: AMBER
      2T). \emph{\bf Middle-left:} Adding super-synthesis to the last
      case improves (u,v) coverage. \emph{\bf Middle-right:} The same
      as before but using 3 telescopes (AMBER today). \emph{\bf
        Bottom-left:} Using 4 telescopes. \emph{\bf Bottom-right:}
      Using 8 telescopes (4 Ats, thin lines and 4 UTs, thick lines).
    }
  }
  \label{fig:uvPlaneFillingSpectralRange}
\end{figure}
\vspace{1cm}

Spectral coverage helps in filling the (u,v) plane simply because
different wavelengths probe different spatial frequencies for a
given baseline:

\begin{equation}
  \rho = \frac{B}{\lambda}
\end{equation}

Thus, using a spectroscopic interferometer allows one to scan a line
in the (u,v) plane with only one snapshot (see
Fig.~\ref{fig:uvPlaneFillingSpectralRange}, top-right). The more
spectral range an instrument has, the more (u,v) plane filling it
gives.

Moreover, spectral coverage also allows one to greatly improve the
number of measures one can access since more phase information is
accessible via differential phase, as explained in detail in
\citet{MillourThesis}.

However, such improvement assumes that the object's shape is
achromatic with regards to wavelength. For example, in the specific
case where the object's shape is different for each wavelength, then
this spectral coverage property does not improve (u,v) coverage.

As a matter of conclusion, combining many telescopes, a long observing
time, and spectral coverage allows one to get a better (u,v) coverage
of an object. Since the number of telescopes is very limited and many
instruments were not able to spectrally analyse the
light until recently, the only way was to increase the exposure
time. Today, both the
spectral coverage and the number of telescopes is increasing (6 telescopes
for CHARA and simultaneous J, H and K bands for AMBER are examples),
which allows one to begin imaging of the targets.

\section{The shape-visibility relation}
\label{sect:shapvis}

Here I add several notations to the previous ones:

\begin{itemize}
\item  $r = \|\overrightarrow{s}\|$,
\item $\rho = \|\overrightarrow{u}\|$,
\item $a, l, L$ are the object's typical dimensions (diameter, FWHM,
  etc. if any), 
\item $R$ is a flux ratio (if any).
\end{itemize}

As seen in previous sections, the measurement of an interferometer is related
to the FT of the brightness distribution of the
object. The goal of this section is to show the
(complex) visibility description and to have some generic laws one can
use to perform a simple interpretation.

\subsection{Generic properties}

First of all, one always has to have in mind that interferometry deals
with FTs. We recall here the generic properties of FTs for
continuous functions:

\begin{description}
\item[\textbullet\  linearity (addition):] $FT[f + g] = FT[f] + FT[g]$,
\item[\textbullet\ translation (shift):] $FT[f(x-x_0,y-y_0)] =
  FT[f](u,v) \times e^{2 i \pi, (u x_0 + v y_0)}$,
\item[\textbullet\ similarity (zoom and shrink):] $FT[f(\alpha x, \beta y)] = \frac{1}{\alpha\beta}FT[f](\frac{u}{\alpha},\frac{v}{\beta})$,
\item[\textbullet\ convolution (``blurring''):] $FT[f \otimes g] =
  FT[f] \times FT[g]$,
\item[\textbullet\ $\infty$ limit (``small'' details):] $FT[f] \stackrel{\infty}{\longmapsto} 0$,
\item[\textbullet\ $0$ limit (``large'' details):] $FT[f] \stackrel{0}{\longmapsto} 1$.
\end{description}

\begin{figure}[htbp]
  \centering
  \includegraphics[width=0.7\textwidth]{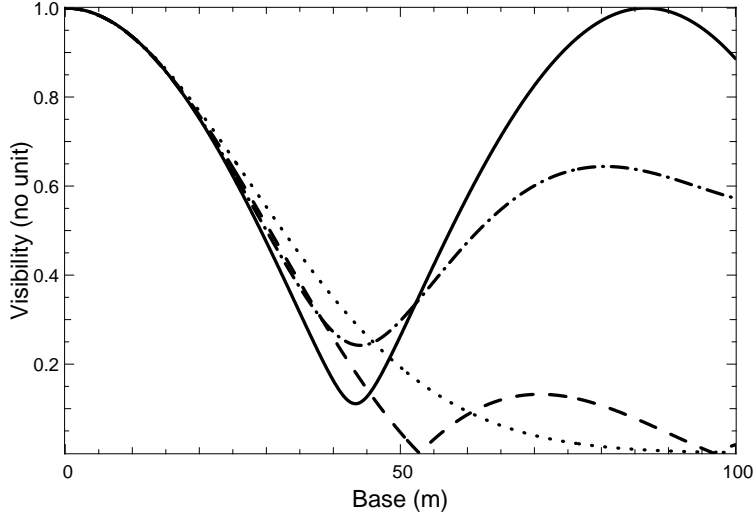}
  \caption[Different models visibilities]{
    \footnotesize{
      Whichever model is used (binary star in solid line, Gaussian disk in
      dotted line, uniform disk in dashed line, or a composite object
      in dash-dotted line), the small baselines do not allow one to
      distinguish it's shape.
    }
  }
  \label{fig:basesVisibility}
\end{figure}

The last two points lead to a specificity of optical long-baseline
stellar interferometry: the best constraints for a model will be given
by a measured visibility of $0.5$. This translates into:

\begin{itemize}
\item For short baselines (i.e. small spatial frequencies), all the
  possible shapes have degenerate visibilities. The visibility modulus
  has a squared dependency with base length \citep[see
  Fig.~\ref{fig:basesVisibility} and the demonstration
  in][]{2003A&A...400..795L} and its phase has a linear
  dependency.
\item For long baselines, all continuous models (i.e., the ones not
  including $\delta$ functions) have visibilities towards
  0. For a given accuracy on the measurements, therefore, no
  discrimination between different models can be done as the object is
  ``over-resolved''.
\end{itemize}

Therefore, to maximise the efficiency of an observation, one has to
know the approximate shape and size of an object in advance (1$^{st}$
guess). The knowledge of an object must then be as high as possible to
maximise the chances of success of an interferometric observation.
Once the observations have been made, and for a given model,
there are many ways to fit the visibilities without changing the
model itself:

\begin{itemize}
\item change the distance (using similarity),
\item change the orientation in the plane of sky (using linearity
  and coordinate transformations),
\item shrink or expand the model on one axis (using similarity), or
\item blur the model (using convolution).
\end{itemize}

The two first points are useful for model fitting as one model image
FT can be scaled and rotated to fit the observed data.

\subsection{Examples}

Here I present typical examples where the visibility function can be
computed easily with an analytical formula. They are the basis of all
further analysis, allowing one to get a first idea of the object's
shape without doing a complete physical modelling.

\paragraph{Point source: double or multiple star}

A centered point source is described in the Fourier plane by a
constant. For many stars, even if the resolving power of
interferometry is much higher than for single-dish telescopes, a good
approximation of their shape can be made, assuming they are
unresolved. The shift and add properties of the FT
give the visibility expression for a multiple star:

\begin{equation}
  V(\overrightarrow{u}) = \frac{\sum_{k=1}^N I_k \cos\left(\frac{\overrightarrow{u} \cdot \overrightarrow{s_k} }{ \lambda}\right)}{\sum_{k=1}^N I_k}
\end{equation}

\vspace{1cm}
\vspace{1cm}

\paragraph{Gaussian disk: extended envelope}

In a first approximation, a Gaussian brightness distribution can
describe several types of envelopes, such as an opaque wind surface around
Wolf-Rayet stars, gaseous or dusty disks around young stars, etc. A
Gaussian brightness distribution has a Gaussian visibility function:

\begin{equation}
  V(\rho) = e^{- \frac{(\pi a \rho)^2}{4 \ln 2}}
\end{equation}

\paragraph{Uniform disk: stellar surface}

In a first attempt, a uniform disk brightness distribution is well
suited to describe a stellar surface, since, in general, the stars
look like a sharp-edged disk, corresponding to the photosphere surface. The
visibility function of a uniform disk is a 1$^{st}$ order Bessel function
divided by $\pi a \rho$: 

\begin{equation}
  V(\rho) = \frac{\mbox{J}_1(\pi a \rho)}{\pi a
    \rho}
\end{equation}

\vspace{1cm}
\vspace{1cm}

\paragraph{Ring and Hankel function: any circular object}

A ring can describe a thin shell around a star. Its expression
depends on the 0$^{th}$ order Bessel function:

\begin{equation}
  V(\rho) = \mbox{J}_0(\pi a \rho)
\end{equation}

It is also very interesting to use the addition property of the FT to
produce the visibility function of any circular object with a given
profile $I(r)$:

\begin{equation}
  V(\rho) = 2 \pi \int_0^\infty I(r) \mbox{J}_0(2 \pi r \rho) r dr
\end{equation}

\paragraph{Other objects}

For other objects, however, no simple tool except FT can be used. The
best thing is to produce an image for the object in each wavelength
of interest and then  produce its FT and compare it with
the visibilities.

In Table~\ref{tab:Visibilities}, one can see a summary of the
different possibilities offered to test simple models.

\vspace{1cm}
\begin{table}[htbp]
  \caption[Simple models and their associated visibilities]{
    \footnotesize{
      Summary table for simple shapes and their corresponding
      complex visibility.
    }
  }
  \centering
  \begin{tabular}{|c|c|c|}
    \hline
    Shape & Brightness distribution & Visibility \\
    \hline
    \hline
    Point source & $\delta(\overrightarrow{s})$ & $1$\\
    \hline
    Binary star & $A \left[ \delta(\overrightarrow{s}) + R
      \delta(\overrightarrow{s} - \overrightarrow{s_0}) \right]$ &
    $\sqrt{\frac{1+R^2+2 R \cos\left(\frac{\overrightarrow{u} \cdot
            \overrightarrow{s_0} }{ \lambda}\right)}{1 + R^2}}$\\
    \hline
    Gauss & $I_0 \sqrt{\frac{4 \ln (2a)}{\pi}} \times e^{- 4 \ln 2
      \frac{r^2}{a^2}}$ & $e^{- \frac{(\pi a
        \rho)^2}{4 \ln 2}}$ \\
    \hline
    Uniform disk & 
    $
    \left\{
      \begin{array}{rl}
        \frac{4}{\pi a^2} & \mbox{if } r < \frac{a}{2}\\
        0 & \mbox{otherwise} \\
      \end{array}
    \right.
    $
    & $\frac{\mbox{J}_1(\pi a \rho)}{\pi a
      \rho}$\\
    \hline
    Ring & $\frac{1}{\pi a} \delta\left(r -
      \frac{a}{2}\right)$ & $\mbox{J}_0(\pi a \rho)$
    \\
    \hline
    Any circular object & $I(r)$ & $2 \pi
    \int_0^\infty I(r) \mbox{J}_0(2 \pi r \rho) r
    dr$ \\
    \hline
    Pixel (image brick)&
    $
    \left\{
      \begin{array}{rl}
        \frac{1}{l L} & \mbox{if } x < l \mbox{ and }  y < L\\
        0 & \mbox{otherwise} \\
      \end{array}
    \right.
    $
    & $\frac{\sin(\pi x l) \sin(\pi y L)}{\pi^2 xy lL}$
    \\
    \hline
  \end{tabular}
  \label{tab:Visibilities}
\end{table}
\vspace{1cm}

\section{Conclusions}

I have reported the main things to know about astronomical optical
interferometry: a short history, the main principles, and the
main properties of the measurements one gets from optical long baseline
interferometry. Using this information, and by doing the practical
exercises associated with this article, a future observer should
be able to prepare observations knowing the real
limitations but also the true possibilities of today's interferometers.

\section*{Acknowledgements}

The author thank the Max-Planck Institut for Radioastronomy for
support and funding through a stipend for the years 2007 and
2008. Thanks also to Gerhard H\"udepohl who allowed use of one of his
aerial photos in the AMBER data reduction software. Part of this work
and figures were made with the
Yorick\footnote{\url{http://yorick.sourceforge.net}} scientific
software and amdlib\footnote{\url{http://amber.obs.ujf-grenoble.fr}, section
data processing} AMBER data reduction software.

% \appendix

% \section{}
% \label{}

% Bibliographic references with the natbib package:
% Parenthetical: \citep{Bai92} produces (Bailyn 1992).
% Textual: \citet{Bai95} produces Bailyn et al. (1995).
% An affix and part of a reference:
% \citep[e.g.][Ch. 2]{Bar76}
% produces (e.g. Barnes et al. 1976, Ch. 2).

% \bibliography{biblio}
% \bibliographystyle{aa}

\end{document}